%% file: arXiv_v4.tex
\documentclass[%
twocolumn,
 amsmath,amssymb,
 aps,
floatfix,
]{revtex4-2}

\usepackage[dvipdfmx]{graphicx}
\usepackage{dcolumn}
\usepackage{bm}
\usepackage{here}
\usepackage{newtxtext,newtxmath}
\usepackage{physics}
\usepackage{xspace}
\usepackage{xcolor}
\usepackage{ulem}
\usepackage{booktabs}
\usepackage{soulutf8}
\usepackage{amsmath}
\usepackage{cancel}
\usepackage{ragged2e}

\usepackage{multirow}

\renewcommand{\Re}{\operatorname{Re}}

\usepackage[unicode=true,colorlinks=true,linkcolor=blue,citecolor=blue,urlcolor=blue,hypertexnames=false]{hyperref}
\usepackage{cleveref}

\begin{document}


\title{Efficient first-principles approach to Gibbs free energy with thermal expansion}

\author{Kota Hashimoto}
 \email{hashimoto.k.d594@m.isct.ac.jp}

\author{Tomonori Tanaka}

\author{Yoshihiro Gohda}
 \email{gohda@mct.isct.ac.jp}

\affiliation{
Department of Materials Science and Engineering, Institute of Science Tokyo, Yokohama 226-8501, Japan
}




\date{\today}

\begin{abstract}
We propose a method to evaluate the Gibbs free energy from constant-volume first-principles phonon calculations.
The volume integral of the pressure is performed by determining the volume and the bulk modulus in equilibrium at finite temperatures, where the pressure and its volume derivative are evaluated utilizing first-principles calculations of the Gr\"{u}neisen parameter without varying the volume.
We validate our method for fcc Al by comparing with the conventional quasiharmonic approximation.
Furthermore, we integrate our method with self-consistent phonon theory and apply it to calculations for bcc Ti, hcp Ti, and tetragonal ZrO$_2$.
We demonstrate the accuracy and computational efficiency of our method by comparing results with those obtained from directly volume-varied self-consistent phonon calculations.
In all cases, our method accurately evaluates the free energy change due to thermal expansion using only constant-volume phonon calculations.

\end{abstract}

\maketitle
\clearpage

\section{Introduction}
The Gibbs free energy is a critical quantity for understanding the thermodynamic properties of materials.
Moreover, it serves as the basis for not only phase stability but also nonequilibrium dynamics, such as the evolution of microstructures \cite{Chen2002-tt,Steinbach2009-rp}.
Due to its importance, countless experiments have been conducted and compiled into databases over the years, in line with the development of the CALPHAD (CALculation of PHAse Diagrams) method.
However, the evaluation of the free energy based solely on experiments has become increasingly difficult with the growing complexity of multicomponent materials.
Therefore, there has been a growing interest in thermodynamics assessment through first-principles calculations, particularly using density functional theory (DFT) \cite{martin_2004}, to address this challenge \cite{liu_thermodynamics_2023}.
Although DFT is valid only at the ground state, finite-temperature properties can be estimated by combining statistical mechanical models with DFT.
Additionally, data assimilation, which integrates numerical and experimental data, allows for the high-precision prediction of materials properties, even when experimental data is limited \cite{Harashima2021-xw,Yamamura2022-kg}.
These high-precision predictions have the potential to accelerate materials development by enhancing the accuracy of free energy calculations.

Various methods exist for incorporating thermodynamic contributions into first-principles free energy assessments: cluster-expansion and variation methods for mixing enthalpy and entropy \cite{kikuchi_theory_1951,van_baal_order-disorder_1973,sanchez_fcc_1978,sanchez_generalized_1984,mohri_short_1985,kikuchi_cvm_1994,Enomoto2023-li}, phonon calculations for vibrational free energy \cite{Dove1993-ce,Nishino2023-gb}, and spin models for magnetic free energy.
Furthermore, interactions between different contributions are crucial for the quantitative evaluation of the free energy \cite{mauger_nonharmonic_2014,tanaka_first-principles_2020,Tanaka2020-tn}.
For an accurate assessment of the free energy over a wide range of temperatures, the contribution of thermal expansion is also important.

Thermal expansion originates primarily from lattice anharmonicity, at least in nonmagnetic materials.
A prominent computational method for evaluating thermal expansion is the quasiharmonic approximation (QHA) \cite{Pavone1993-or,Karki2000-xl,Mounet2005-go,Ritz2018-iy,Togo2010-cz,Huang2016-gd,Abraham2018-es,Allen2015-jl,Allen2020-my,Bakare2022-ze,Masuki2022-ev,Masuki2023-fy,Rostami2024-ea}.
Within the QHA, the equilibrium volume at a finite temperature is determined by performing harmonic phonon calculations on structural models with varying volumes.
However, the QHA is not applicable to high-temperature phases that exhibit lattice instability at low temperatures, due to imaginary frequencies.
To restore lattice stability at finite temperatures, it is necessary to consider the anharmonic effects of phonons.
Several approaches exist to incorporate anharmonicity, such as temperature-dependent effective potentials \cite{hellman_temperature_2013,hellman_temperature-dependent_2013,romero_thermal_2015}, piecewise polynomial potential partitioning \cite{kadkhodaei_free_2017, kadkhodaei_software_2020}, self-consistent ab initio lattice dynamics \cite{souvatzis_entropy_2008}, and self-consistent phonon (SCPh) theory \cite{Tadano2015-hf, Tadano2018-jk}.
Unfortunately, in any of these approaches, the direct modeling of thermal expansion, as in the QHA, is computationally expensive because these methods require substantial resources even for a single-volume calculation.
This poses a crucial challenge in the systematic evaluation of free energies for materials with diverse compositions.
Previous studies have highlighted the significance of thermal expansion, emphasizing its impact on phase stability and phase-transition temperatures \cite{Quong1997-qw,Van_de_Walle2002-ap,Zhang2013-op,Wang2010-ug,Togo2015-pd,Ritz2019-gp,Masuki2022-cp,Masuki2023-fy}.
Therefore, an efficient approach for evaluating the effects of thermal expansion is in demand.

In this study, we present an alternative approach that evaluates the change in the free energy due to thermal expansion from phonon calculations at a single volume.
This approach eliminates the need for computationally demanding calculations involving many different volumes of supercells.
Our approach is based on the combination of the vibrational pressure and bulk modulus via Gr\"{u}neisen theory \cite{Gruneisen1912-he}.
From phonon calculations at a single volume, we obtain the equilibrium volume and equilibrium bulk modulus, which serve as parameters in the second-order Birch-Murnaghan equation of state at any temperature.
Using these parameters, the change in the free energy due to thermal expansion is calculated by integrating pressure with respect to volume.
We perform first-principles calculations of the Gibbs free energy and the coefficient of linear thermal expansion for fcc Al, which is a system with large thermal expansion coefficient, to validate our approach.
Furthermore, to demonstrate the accuracy and computational efficiency of our approach, we integrate it with SCPh theory and perform calculations of bcc and hcp Ti, as well as tetragonal ZrO$_2$, and compare the results with those obtained from directly volume-varied SCPh calculations.

\section{Theory}
The Gibbs free energy $G(T, P)$, where $T$ is the temperature and $P$ is the external pressure, is obtained via a Legendre transformation as
\begin{align}
  G(T,P)&=\underset{V} {\operatorname{min}} \left[F(T,V)+PV\right]\nonumber\\
  &=F(T,V_{\rm eq}(T,P))+PV_{\rm eq}(T,P),
  \label{gibbs}
\end{align}
where $F$ is the Helmholtz free energy, expressed as a function of $T$ and the volume $V$ \cite{Togo2015-pd}, and $V_{\rm eq}$ is the equilibrium volume at $T$ and $P$. 
The mathematical notation of $\min_x[f(x)]$ means finding the minimum value of the function $f(x)$ by changing the variable $x$.
Therefore, $V_{\rm eq}$ can be obtained as
\begin{align}
  \label{veq}
  V_{\rm eq}(T,P)=\underset{V} {\operatorname{argmin}} \left[F(T,V)+PV\right],
\end{align}
where $\operatorname{argmin}_x[f(x)]$ means the value of variable $x$ at which the function $f(x)$ is minimized. According to Eq.~(\ref{gibbs}), evaluating $G(T, P)$ typically requires Helmholtz free energies over a range of volumes, since the minimization is performed with respect to volume. Instead, we introduce a practical scheme that reconstructs $G(T, P)$ using only the Helmholtz free energy at a fixed reference volume $V_0$, i.e., $F(T, V_0)$, aided by an equation of state.

\subsection{Volume integral of pressure (VIP) method}
Our objective is to determine the Gibbs free energy $G(T,\{N_i\},P)$ of a single phase, where $\{N_i\}$ is a set of numbers of atoms with the element index $i$.
In the following discussion, we consider closed systems.
Therefore, we do not write $\{N_i\}$ as a variable explicitly.
The Gibbs free energy $G(T,P)$ can be written as
\begin{align}
  G(T,P)&=F(T,V_{\rm{eq}}(T, P))+PV_{\rm{eq}}(T, P)\nonumber\\
  &=F(T,V_0)-\int_{V_0}^{V_{\rm{eq}}}P(T,V)dV+PV_{\rm{eq}}(T, P),
 \label{gibbs_trans}
\end{align}
where $P(T, V)$ is the pressure as a function of $T$ and $V$.

To calculate the integral term in Eq.~(\ref{gibbs_trans}), we propose a method named as the volume integral of pressure (VIP) method.
Here, we use the second-order Birch-Murnaghan equation of state (EOS) \cite{Birch1938-ic,Birch1947-yw,Katsura2019-pe}
\begin{align}
P(T,V)&=\frac{3B_{\rm{eq}}}{2}\left[\left(\frac{V_{\rm{eq}}}{V}\right)^{\frac{7}{3}}-\left(\frac{V_{\rm{eq}}}{V}\right)^{\frac{5}{3}}\right],
\label{eos}
\end{align}
where $B_{\rm{eq}}$ is the bulk modulus at $V_{\rm{eq}}$.
The parameters $B_{\rm{eq}}$ and $V_{\rm{eq}}$ can be determined by evaluating $P(T, V_0)$ and its first derivative
\begin{align}
\left.\left(\frac{\partial{P}}{\partial{V}}\right)_T\right|_{V_0}
=\left.\frac{3B_{\rm{eq}}}{2}\left(\frac{\partial}{\partial{V}}\right)_T\left[\left(\frac{V_{\rm{eq}}}{V}\right)^{\frac{7}{3}}-\left(\frac{V_{\rm{eq}}}{V}\right)^{\frac{5}{3}}\right]\right|_{V_0},
\label{eos_derivative}
\end{align}
at $V_0$.
The left-hand sides of Eqs.~(\ref{eos}) and (\ref{eos_derivative}) are numerically calculated from first principles.
Thus, $B_{\rm eq}$ and $V_{\rm eq}$ are determined by solving the set of equations.
For nonmagnetic materials, the Helmholtz free energy $F(T,V)$ within the Born-Oppenheimer approximation can be divided into the following contributions \cite{Fletcher1979-gf}
\begin{align}
  F(T,V)=E(T\!=\!0,V)+F_{\mathrm{ele}}(T,V)+F_{\mathrm{vib}}(T,V),
  \label{helmholtz}
\end{align}
where $E(T\!=\!0,V)$ is the internal energy at zero temperature except for the zero-point vibrational energy, $F_{\rm{ele}}$ is the electronic free energy defined as a change from the zero-temperature case, and $F_{\rm{vib}}$ is the vibrational free energy.
For a given atomic configuration, $E(T\!=\!0,V)$ is calculated as the ground-state total energy from first principles.
The electronic contributions to the free energy can be calculated within the finite-temperature DFT \cite{Grabowski2007-kw,Kormann2008-rq}.
From the relation
\begin{align}
  P(T,V)
  =-\left(\frac{\partial{F}}{\partial{V}}\right)_T,
  \label{pressure}
\end{align}
the pressure and its volume derivative for each term in Eq.~(\ref{helmholtz}) are obtained.
Since computational costs of $E(T\!=\!0,V)+F_{\mathrm{ele}}(T,V)$ are in general modest, a set of this quantity for different volumes can be used to calculate the contributions to the pressure $P_{\rm ele}$ and its volume derivative $(\partial P_{\rm ele}/\partial V)_T$.
In contrast, the key point of the VIP method is that we evaluate the vibrational contributions to Eqs.~(\ref{eos}) and (\ref{eos_derivative}) from first-principles phonon calculations at single volume $V_0$, as explained in the following subsection.

\subsection{Vibrational pressure and its volume derivative}
\label{vip_details}
We derive analytical formulae for $P_{\rm{{vib}}}(T,V)$ and its volume derivative, to evaluate the left hand sides in Eqs.~(\ref{eos}) and (\ref{eos_derivative}).
First, we express the vibrational pressure $P_{\rm vib}(T,V)$ in terms of the vibrational free energy $F_{\rm vib}$ as follows:
\begin{align}
  \label{F_vib}
  F_{\rm{vib}}(T,V)
  &=k_{\rm{B}}T\sum_{\boldsymbol{q},j}\log\left[2\sinh{\frac{\hbar\omega_{\boldsymbol{q}j}(V)}{2k_{\rm{B}}T}}\right],\\
  P_{\rm{vib}}(T,V)
  \label{P_vib}
  &=-\left(\frac{\partial F_{\rm vib}}{\partial V}\right)_T\nonumber\\
  &=-\sum_{\boldsymbol{q},j}\left(\frac{\partial{F_{\rm{vib}}}}{\partial{\omega_{\boldsymbol{q}j}}}\right)_T\left(\frac{\partial{\omega_{\boldsymbol{q}j}}}{\partial{V}}\right)_T\nonumber\\
  &=\frac{\hbar}{V}\sum_{\boldsymbol{q},j}\gamma_{\boldsymbol{q}j}(V)\omega_{\boldsymbol{q}j}(V)\left[n_{\rm{B}}(T,\hbar\omega_{\boldsymbol{q}j}(V))+\frac{1}{2}\right],
\end{align}
\begin{align}
  \label{dF_domega}
  \left(\frac{\partial{F_{\rm{vib}}}}{\partial{\omega_{\boldsymbol{q}j}}}\right)_T
  =\frac{\hbar}{2}\tanh{\frac{\hbar\omega_{\boldsymbol{q}j}(V)}{2k_{\rm{B}}T}}
  =\hbar\left[n_{\rm{B}}(T,\hbar\omega_{\boldsymbol{q}j}(V))+\frac{1}{2}\right],
\end{align}
where $\omega_{\boldsymbol{q}j}(V)$ is the volume-dependent frequency of the $j$th phonon with the wave vector $\boldsymbol{q}$, $k_{\rm{B}}$ is the Boltzmann constant, $n_{\rm{B}}(T,\hbar\omega_{\boldsymbol{q}j}(V))$ is the Bose-Einstein distribution function with zero chemical potential, and $\gamma_{\boldsymbol{q}j}(V)$ is the mode Gr\"{u}neisen parameter \cite{Ritz2019-gp,Gruneisen1912-he,Dove1993-ce,Ashcroft2011} as follows:
\begin{align}
  \gamma_{\boldsymbol{q}j}(V)
  &=-\frac{V}{\omega_{\boldsymbol{q}j}(V)}\frac{\partial{\omega_{\boldsymbol{q}j}}(V)}{\partial{V}}\\
  &=-\frac{(\boldsymbol{e}_{\boldsymbol{q}j}^*)^{\rm{T}}\delta{D}(\boldsymbol{q})\boldsymbol{e}_{\boldsymbol{q}j}}{6\omega_{\boldsymbol{q}j}^2(V)},
  \label{gruneisen}
\end{align}
where $\boldsymbol{e}_{\boldsymbol{q}j}$ is the polarization vector and $\delta{D}(\boldsymbol{q})$ is the variation of the dynamical matrix due to a volume change.
Appendix \ref{Gruneisen_parameter_definitions} contains the details of the definitions.
The derivative of $P_{\rm{vib}}$ with respect to the volume can be written as
\begin{align}
\label{dPdV}
\left(\frac{\partial{P_{\rm{vib}}}}{\partial{V}}\right)_T
=&-\sum_{\boldsymbol{q},j}\left(\frac{\partial}{\partial{V}}\right)_T
\left[\left(\frac{\partial{F_{\rm{vib}}}}{\partial{\omega_{\boldsymbol{q}j}}}\right)_T\left(\frac{\partial{\omega_{\boldsymbol{q}j}}}{\partial{V}}\right)_T\right]\nonumber\\
=&-\sum_{\boldsymbol{q},j}\left[\left(\frac{\partial{\omega_{\boldsymbol{q}j}}}{\partial{V}}\right)_{T}\left(\frac{\partial}{\partial{V}}\right)_{T}\left(\frac{\partial{F_{\rm{vib}}}}{\partial{\omega_{\boldsymbol{q}j}}}\right)_T\right.\nonumber\\
&+\left.\left(\frac{\partial{F_{\rm{vib}}}}{\partial{\omega_{\boldsymbol{q}j}}}\right)_T\left(\frac{\partial}{\partial{V}}\right)_{T}\left(\frac{\partial{\omega_{\boldsymbol{q}j}}}{\partial{V}}\right)_T\right]\nonumber\\
=&-\sum_{\boldsymbol{q},j}\left[\frac{\gamma_{\boldsymbol{q}j}^2{\omega_{\boldsymbol{q}j}}^2}{V^2}\left(\frac{\partial^2{F_{\rm{vib}}}}{\partial{\omega_{\boldsymbol{q}j}}^2}\right)_T\right.\nonumber\\
&+\left.\left(\frac{\partial{F_{\rm{vib}}}}{\partial{\omega_{\boldsymbol{q}j}}}\right)_T\left(
  \frac{\gamma_{\boldsymbol{q}j}\omega_{\boldsymbol{q}j}}{V^2}
  -\frac{1}{V}\left(\frac{\partial{(\gamma_{\boldsymbol{q}j}}\omega_{\boldsymbol{q}j})}{\partial{V}}\right)_T
  \right)\right],\\
\left(\frac{\partial^2{F_{\rm vib}}}{\partial{\omega_{\boldsymbol{q}j}}^2}\right)_T
=&-\frac{\hbar^{2}}{4k_{\rm{B}}T\sinh^{2}{\frac{\hbar\omega_{\boldsymbol{q}j}(V)}{2k_{\rm{B}}T}}}.
\end{align}

\subsection{Volume dependence of parameters}
\begin{figure}[b]
  \begin{center}
  \includegraphics[width=1.0\linewidth]{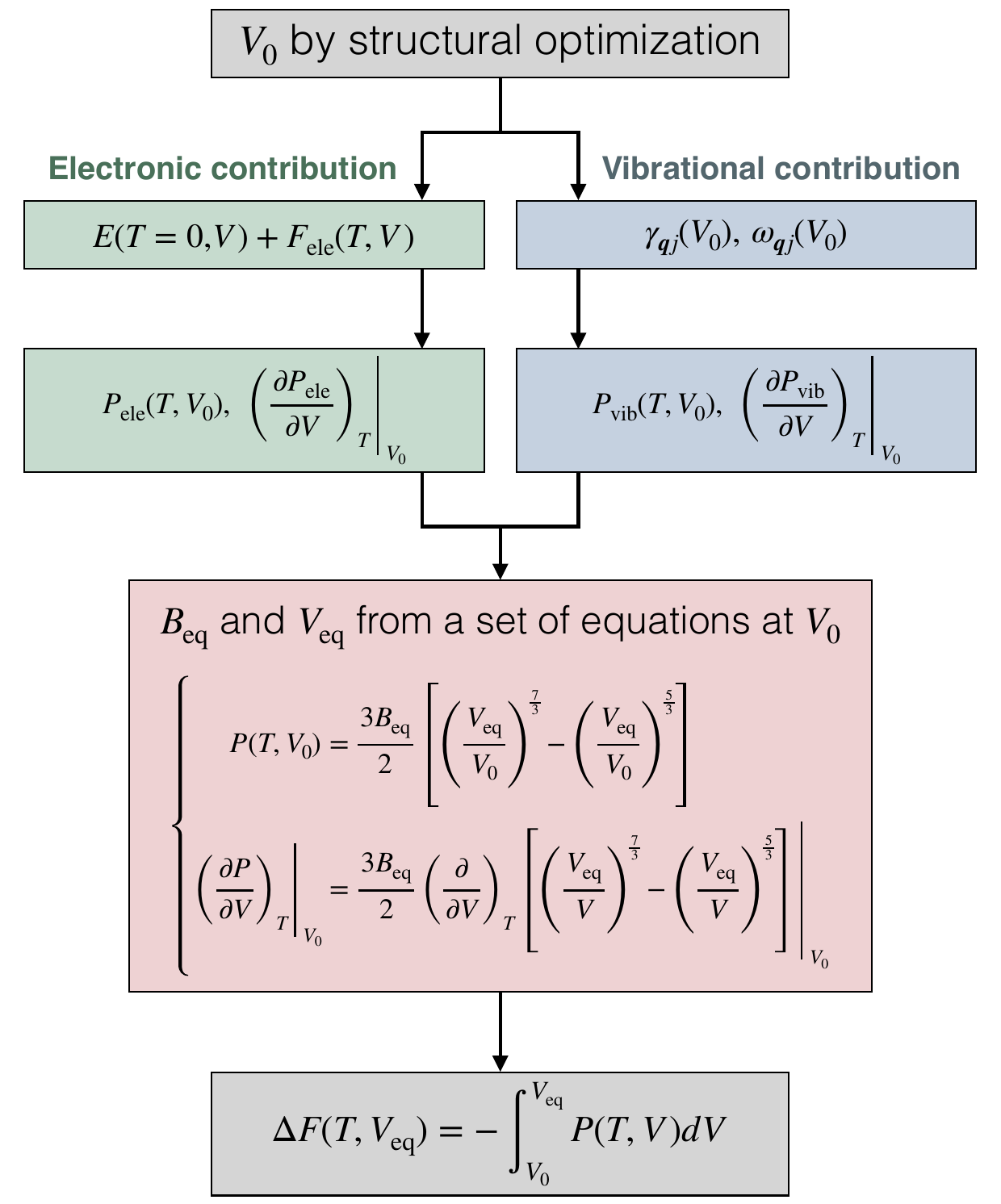}
  \caption{\label{fig:VIP_scheme}
  Flowchart of the VIP method.
  Pressures and their volume derivatives are calculated for each electronic and vibrational contribution at $V_0$.
  The equilibrium volume $V_{\rm eq}$ and bulk modulus $B_{\rm eq}$ at each temperature are subsequently determined by solving a set of equations.
  Finally, the Helmholtz free energy change due to thermal expansion is obtained by integrating the pressure, described by the second-order Birch-Murnaghan equation of state, from $V_0$ to $V_{\rm eq}$.
  }
  \end{center}
\end{figure}

If the contribution of the term containing
$\left(\left(\partial{\gamma_{\boldsymbol{q}j}}\omega_{\boldsymbol qj}\right)/\partial{V}\right)_T$
in Eq.~(\ref{dPdV}) is minor, we can evaluate the Gibbs free energy without phonon calculations at various volumes.
To test this assumption, we redefine Eq.~(\ref{dPdV}) as
\begin{align}
  \label{total_B}
  \left(\frac{\partial{P_{\rm{vib}}}}{\partial{V}}\right)_T=-\frac{1}{V}\left(B_1+B_2\right),
\end{align}
where $B_1$ and $B_2$ are defined as follows:
\begin{align}
  \label{B_1}
  B_1&=\frac{1}{V}\sum_{\boldsymbol{q},j}\left(\gamma_{\boldsymbol{q}j}^2\omega_{\boldsymbol{q}j}^2\left(\frac{\partial^2{F_{\rm{vib}}}}{\partial{\omega_{\boldsymbol{q}j}}^2}\right)_T
  +\gamma_{\boldsymbol{q}j}\omega_{\boldsymbol{q}j}\left(\frac{\partial{F_{\rm{vib}}}}{\partial{\omega_{\boldsymbol{q}j}}}\right)_T\right),\\
  \label{B_2}
  B_2&=-\sum_{\boldsymbol{q},j}\left(\frac{\partial{(\gamma_{\boldsymbol{q}j}}\omega_{\boldsymbol{q}j})}{\partial{V}}\right)_T\left(\frac{\partial{F_{\rm{vib}}}}{\partial{\omega_{\boldsymbol{q}j}}}\right)_T.
\end{align}
Note that Eq.~(\ref{total_B}) is directly related to the bulk modulus as
\begin{align}
  B=-V\left(\frac{\partial{P}}{\partial{V}}\right)_T.
  \label{bulk_modulus}
\end{align}
To evaluate the impact of the $B_2$ on the Gibbs free energy, we performed calculations for fcc Al, diamond Si, bcc Ti, hcp Ti, and tetragonal ZrO$_2$.
These calculations explicitly varied the volume to assess the influence of the $B_2$ on the Gibbs free energy.
Our results show that the contribution of the $B_2$ is negligible in all cases. 
These numerical results are presented in Sec. S2 of the Supplemental Material \cite{Supplemental}.
Therefore, Eq.~(\ref{dPdV}) can be approximated as
\begin{align}
  \label{dPdV_approx}
  \left(\frac{\partial{P_{\rm{vib}}}}{\partial{V}}\right)_T
  \thickapprox \frac{-1}{V^2}\sum_{\boldsymbol{q},j}\left[\gamma_{\boldsymbol{q}j}^2{\omega_{\boldsymbol{q}j}}^2\left(\frac{\partial^2{F_{\rm{vib}}}}{\partial{\omega_{\boldsymbol{q}j}}^2}\right)_T\right.+\gamma_{\boldsymbol{q}j}\omega_{\boldsymbol{q}j}\left.\left(\frac{\partial{F_{\rm{vib}}}}{\partial{\omega_{\boldsymbol{q}j}}}\right)_T\right].
  \end{align}
From the above discussions, the flowchart of the VIP method is as illustrated in Fig~\ref{fig:VIP_scheme}.
The true strength of the VIP method lies in its seamless integration with SCPh calculations.
During the fitting process to obtain fourth-order force constants for SCPh calculations, the third-order force constants are determined simultaneously, thereby eliminating any additional computational costs \cite{Tadano2015-hf}.

\subsection{Computational details}
We performed first-principles calculations based on density functional theory within the projector augmented wave (PAW) method \cite{Blochl1994-ek} as implemented in the Vienna $ab\ initio$ simulation package (VASP) \cite{Kresse1996-en,Kresse1999-rv}.
The generalized gradient approximation (GGA) parameterized by Perdew, Burke, and Ernzerhof (PBE) \cite{Perdew1996-aq} was employed for the exchange-correlation functional.
Supercells of 3$\times$3$\times$3 (108 atoms) for fcc Al, and 4$\times$4$\times$4 (128 atoms) for both hcp and bcc Ti were employed.
The cutoff energy was set to 313 eV and 232 eV for Al and Ti, respectively.
A $\Gamma$-centered $6\times6\times6$ $k$-point grid was used for Al, while $\Gamma$-centered $5\times5\times5$ and $7\times7\times4$ grids were used for bcc and hcp Ti, respectively.
For tetragonal ZrO$_2$, we used a $3\times3\times2$ supercell with 108 atoms and a Monkhorst-Pack $k$-point grid of $6\times6\times4$.
The cutoff energy was set to 520 eV and the exchange-correlation functional was treated within the PBEsol \cite{Perdew2008-ta} according to Refs.~\cite{Tolborg2023-ep,Verdi2021-jl,Delarmelina2020-wk}.
Interatomic force constants (IFCs) and phonon frequencies were calculated using the ALAMODE package \cite{Tadano2014-fv}.
The harmonic IFCs were calculated using the so-called direct method with atomic displacements of 0.01 \AA{}.
To calculate mode Gr\"{u}neisen parameters, third-order IFCs were estimated using the least absolute shrinkage and selection operator (LASSO) technique from randomly displaced structures, as implemented in the ALAMODE package \cite{Tadano2014-fv}.
For hcp Ti, we explicitly optimized the $c/a$ ratio at each temperature.
Our calculations reveal that the $c/a$ ratio changes very small, with a maximum variation of only about 0.3 \% over the studied temperature range.
This small change is consistent with the experimental measurement \cite{Touloukian1975-tv} and the previous computational study \cite{Jung2023-yz} and it supports our approximation that thermal expansion along the $a$ and $c$ axes is nearly identical.
The $c/a$ ratio is presented in Sec. S1 of the Supplemental Material \cite{Supplemental}.
For bcc and hcp Ti, we employed $ab\ initio$ molecular dynamics (AIMD) at 1155 K to generate atomic configurations at finite temperature. 
Moreover, harmonic IFCs of hcp Ti were extracted from atomic configurations at finite temperature generated by AIMD at 300 K.
In the estimations of IFCs, we considered anharmonic terms up to the sixth and fourth order for bcc and hcp Ti, respectively.
In addition, for tetragonal ZrO$_2$, we employed AIMD to generate random atomic configurations at 1000 K and performed SCPh calculations including fourth order IFCs.
We calculated the Born effective charges and the dielectric tensor of tetragonal ZrO$_2$ using density functional perturbation theory and obtained values of $\varepsilon^{\infty}_{xx}=\varepsilon^{\infty}_{yy}=5.79$, $\varepsilon^{\infty}_{zz}=5.19$, $Z_{xx}^{*}(\rm Zr)=Z_{yy}^{*}(\rm Zr)=5.76$, $Z_{zz}^{*}(\rm Zr)=4.98$, $Z_{xx}^{*}(\rm O)=-3.62$, $Z_{yy}^{*}(\rm O)=-2.13$, $Z_{zz}^{*}(\rm O)=-2.49$, which agree well with the previous computational result \cite{Zhang2018-kh}.
The vibrational free energy based on SCPh theory is represented in Eq.~(\ref{F_vib}) with an additional correction term \cite{Tadano2014-fv,Oba2019-ja}. 
However, this correction term was disregarded when calculating the vibrational pressure and its volume derivative because its contribution is small compared with the harmonic term.
Therefore, the vibrational frequency $\omega_{\boldsymbol qj}(V)$ was simply replaced with $\Omega_{\boldsymbol qj}(T,V)$.
The electronic free energy calculated by using the density of state (DOS) $D(V,\varepsilon)$ and the Fermi-Dirac distribution function $f(T, \varepsilon)$ as follows \cite{Zhang2017-cj}:
\begin{align}
  \label{F_ele}
  F_{\rm ele}(T,V) &= E_{\rm ele}(T,V)-T S_{\rm ele}(T,V),\\
  E_{\rm ele}(T,V) &= \int_{-\infty}^{\infty} \varepsilon D(V,\varepsilon)f(T,\varepsilon) d\varepsilon,\\
  S_{\rm ele}(T,V) &= k_{\rm B} \int_{-\infty}^{\infty}D(V,\varepsilon)s(T,\varepsilon) d\varepsilon,
\end{align}
where
\begin{align}
  \label{entropy}
  s(T,\varepsilon)=-\left[f(T,\varepsilon)\log f(T,\varepsilon)+(1-f(T,\varepsilon))\log(1-f(T,\varepsilon))\right],
\end{align}
\begin{align}
  \label{fermi_dirac}
  f(T,\varepsilon)=\left[1+\exp\left(\frac{\varepsilon-\varepsilon_{\rm F}(T)}{k_{\rm B}T}\right)\right]^{-1},
\end{align}
and $\varepsilon_{\rm F}(T)$ is the Fermi level.
The $D(V,\varepsilon)$ in Eq.~(\ref{F_ele}) is calculated from the ground-state total energy first-principles calculations.
The electronic pressure $P_{\rm ele}$ and its volume derivative $(\partial P_{\rm ele}/\partial V)_T$ were calculated using the second-order Birch-Murnaghan EOS for convenience.
Note that the order of the Birch-Murnaghan EOS has almost no effect on the determination of $P_{\rm ele}$ and $(\partial P_{\rm ele}/\partial V)_T$.

In the calculations for bcc and hcp Ti, to approximately account for electron-phonon coupling effects,
we also consider the case of replacing the electronic free energy with that for the atomic configurations with vibrational displacements $F_{\rm ele}^{\rm disp}(T,V)$ as follows~\cite{Zhang2017-cj}:
\begin{align}
  F(T,V)=E(T=0,V)+F_{\rm ele}^{\rm disp}(T,V)+F_{\rm vib}(T,V),
  \label{helmholtz_disp}
\end{align}
where
\begin{align}
  \label{F_ele_disp}
  F_{\rm ele}^{\rm disp}(T,V) &= E_{\rm ele}^{\rm disp}(T,V)-T S_{\rm ele}^{\rm disp}(T,V),\\
  E_{\rm ele}^{\rm disp}(T,V) &= \int_{-\infty}^{\infty} \varepsilon D'(T,V,\varepsilon) f(T,\varepsilon)d\varepsilon,\\
  S_{\rm ele}^{\rm disp}(T,V) &= k_{\rm B} \int_{-\infty}^{\infty}D'(T,V,\varepsilon)s(T,\varepsilon) d\varepsilon,
\end{align}
and $D'(T,V,\varepsilon)$ is the temperature-dependent DOS modulated by the vibrational atomic displacements.
In this case, we generate vibrational atomic displacement at finite temperature \cite{Dove1993-ce} using the ALAMODE package \cite{Tadano2014-fv} as follows:
\begin{align}
  \label{vibrational_displacement}
  \boldsymbol u(l\kappa)
  =\frac{1}{\sqrt{NM_{\kappa}}}\sum_{\boldsymbol{q}j}Q_{\boldsymbol{q}j}\Re\left[\boldsymbol e_{\boldsymbol qj}(\kappa)e^{i \boldsymbol q \cdot \boldsymbol r}\right],
\end{align}
where $N$ is the number of atoms in the primitive cell, $M$ is the mass of the $\kappa$ atoms, $e_{q}$ is the polarization vector, and the normal mode amplitude $Q_{\boldsymbol qj}$ are randomly sampled from Gaussian distribution $P\left[Q_{\boldsymbol qj}\right]$ \cite{Togo2023-en,feynman2018statistical} as
\begin{align}
  \label{gaussian}
  P\left[Q_{\boldsymbol qj}\right]
  =\frac{1}{\sqrt{2 \pi \sigma_{\boldsymbol qj}^2}} \exp \left[-\frac{Q_{\boldsymbol qj}^2}{2 \sigma_{\boldsymbol qj}^2}\right],
\end{align}
where $\sigma_{\boldsymbol qj}^2$ is the variance as follows:
\begin{align}
  \label{variance}
  \sigma_{\boldsymbol qj}^2
  =\frac{\hbar}{2\omega_{\boldsymbol qj}}\coth \frac{\hbar \omega_{\boldsymbol qj}}{2 k_{\rm B}T}
  =\frac{\hbar}{\omega_{\boldsymbol qj}}\left(n_{\rm B}(T, \hbar\omega_{\boldsymbol qj})+\frac12\right).
\end{align}
We confirmed that quantitative differences between Helmholtz free energies evaluated by Eqs.~(\ref{helmholtz}) and (\ref{helmholtz_disp}) are negligible for fcc Al due to strong itineracy of valence electrons.
Thus, we used Eq.~(\ref{helmholtz}) for fcc Al, whereas Eq.~(\ref{helmholtz_disp}) were used for bcc and hcp Ti at $V_0$.
For tetragonal ZrO$_2$, Eq.~(\ref{helmholtz}) was employed since the electronic contribution is negligible due to its insulating nature.

\section{Applications}
We apply the VIP method to fcc Al, which is a system with a large coefficient of linear thermal expansion (CLTE), as a validation of our method.
To further illustrate the effectiveness and computational efficiency of the VIP method, we combine it with SCPh theory to perform calculations on bcc and hcp Ti, as well as tetragonal ZrO$_2$, and then compare these results with those obtained from directly volume-varied SCPh calculations.
We also determine the phase transition point of Ti by Gibbs free energy $G(T,P)$ calculated using the VIP method.
In the following calculation we considered zero external pressure ($P=0$), and $V_0$ in Eq.~(\ref{gibbs_trans}) is chosen as the volume obtained from structural optimization at zero temperature.
This selection serves as the reasonable and conventional starting reference point for practical calculations. The sensitivities of the VIP method to the selection of reference volume for fcc Al and Ti are detailed in Sec. S4 of the Supplemental Material \cite{Supplemental}.

\subsection{fcc Al}
\begin{figure}[H]
  \begin{center}
  \includegraphics[width=1.0\linewidth]{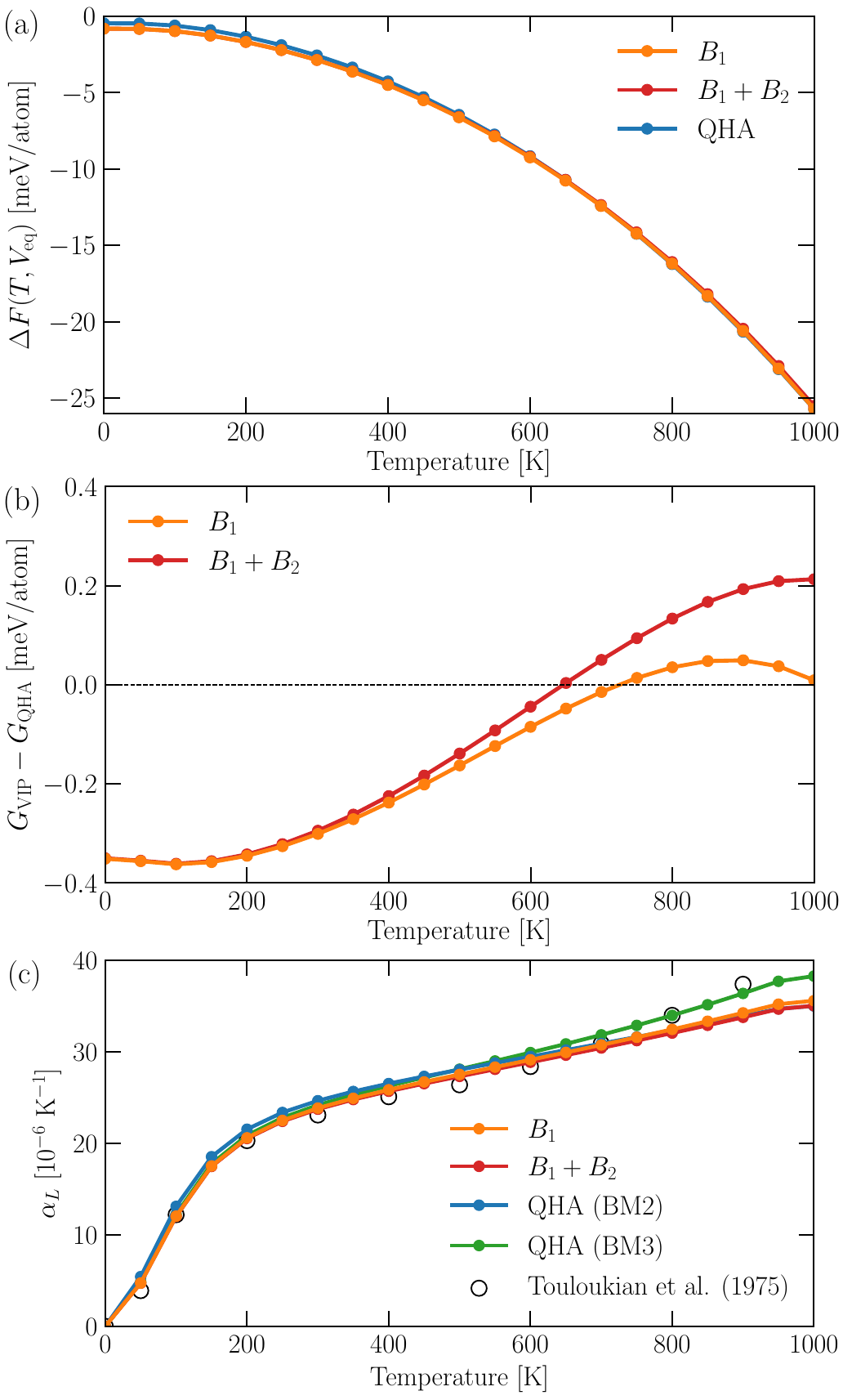}
    \caption{\label{fig:Al_vip_qha_comparison}
    Comparison between the VIP method and the QHA in terms of the free energy and the CLTE for fcc Al over the temperature range from 0 to 1000 K.
    (a) The change in the Helmholtz free energy due to thermal expansion $\Delta F=-\int_{V_0}^{V_{\rm eq}} P(T,V)dV$ in Eq.~(\ref{gibbs_trans}).
    (b) The difference in the Gibbs free energy between the VIP method and the QHA.
    (c) The CLTE $\alpha_L = 1/3 \alpha_V = 1/(3 V_{\rm eq}) \left({dV_{\rm eq}}/{dT}\right)$, where $\alpha_V$ is the coefficient of volumetric thermal expansion and $V_{\rm eq}$ is the equilibrium volume at each temperature.
    The orange line is the standard VIP result ($B_1$ only, Eq.~(\ref{B_1})); the red line is a reference that also includes $B_2$ ($B_1+B_2$, Eqs.~(\ref{B_1}) and (\ref{B_2})).
    The blue and green lines represent the CLTE obtained by the QHA fitted by using the second-order Birch-Murnaghan EOS (BM2) and the third-order Birch-Murnaghan EOS (BM3), respectively.
    The maximum deviation of the CLTE between the VIP method and the QHA, both using BM2, is less than 2\% at 1000 K.
    By comparison, the CLTEs within the QHA derived from the BM2 and BM3 deviate from each other by roughly 10 \% at the same temperature.
    The open circles represent the experimental CLTE by Touloukian et al. \cite{Touloukian1975-tv}.
  }
\end{center}
\end{figure}

Figure~\ref{fig:Al_vip_qha_comparison} shows the comparison between the VIP method and the QHA in terms of the free energy and the CLTE for fcc Al over the temperature range from 0 to 1000 K. 
The $B_2$ term in Eq.~(\ref{B_2}), which arises from the volume dependence of the Gr\"{u}neisen parameters, was evaluated by directly varying the volume as a part of the validation of the VIP method.
We first evaluate the change in the Helmholtz free energy $\Delta F$ due to thermal expansion as shown in Fig.~\ref{fig:Al_vip_qha_comparison}(a).
The change in the Helmholtz free energy is about $-25$ meV/atom at 1000 K.
This indicates that the contribution of thermal expansion to the free energy is significant at high temperatures.
We also compared the Gibbs free energy calculated by the VIP method with that obtained by the QHA, as shown in Fig.~\ref{fig:Al_vip_qha_comparison}(b).
The negligible difference between $B_1+B_2$ and $B_1$ suggests that we can accurately predict the change in the Helmholtz free energy without incorporating $B_2$ into the calculation.
Our method can calculate the Gibbs free energy within $0.5$ meV/atom of the QHA up to 1000 K, even for fcc Al, where the contribution of thermal expansion is significant.
We also calculated the CLTE for fcc Al, as shown in Fig.~\ref{fig:Al_vip_qha_comparison}(c).
The difference in the CLTE between calculations with and without considering $B_2$ is negligible.
The VIP results show a good agreement at lower temperatures,~ up to 600 K,~ with ~a ~deviation of less than ~2\%.
\noindent 
However, a discrepancy exists between the results of the VIP method and the experimental data \cite{Touloukian1975-tv} at high temperatures around 1000 K.
This discrepancy is attributed to the accuracy of EOS; results obtained using the third-order Birch-Murnaghan EOS within the QHA shown as the green line in Fig.~\ref{fig:Al_vip_qha_comparison}(c) significantly reduce the deviation from the experimental data.
Although the error arises from the order of the EOS, we emphasize that even in fcc Al — which has a large coefficient of thermal expansion and is therefore highly sensitive to differences in EOS accuracy — this level of error still allows for an adequate description of the coefficient of thermal expansion.
We therefore expect that the VIP method will accurately evaluate the CLTE of many other systems with smaller thermal expansion than fcc Al.
The details regarding the accuracy of the second-order Birch-Murnaghan EOS for fcc Al are provided in Sec. S3 of the Supplemental Material \cite{Supplemental}.
Moreover, similar discussions for the VIP method applied to diamond Si are available in Sec. S4 in the Supplemental Material \cite{Supplemental}.

\begin{figure}[b]
  \begin{center}
    \includegraphics[width=1.0\linewidth]{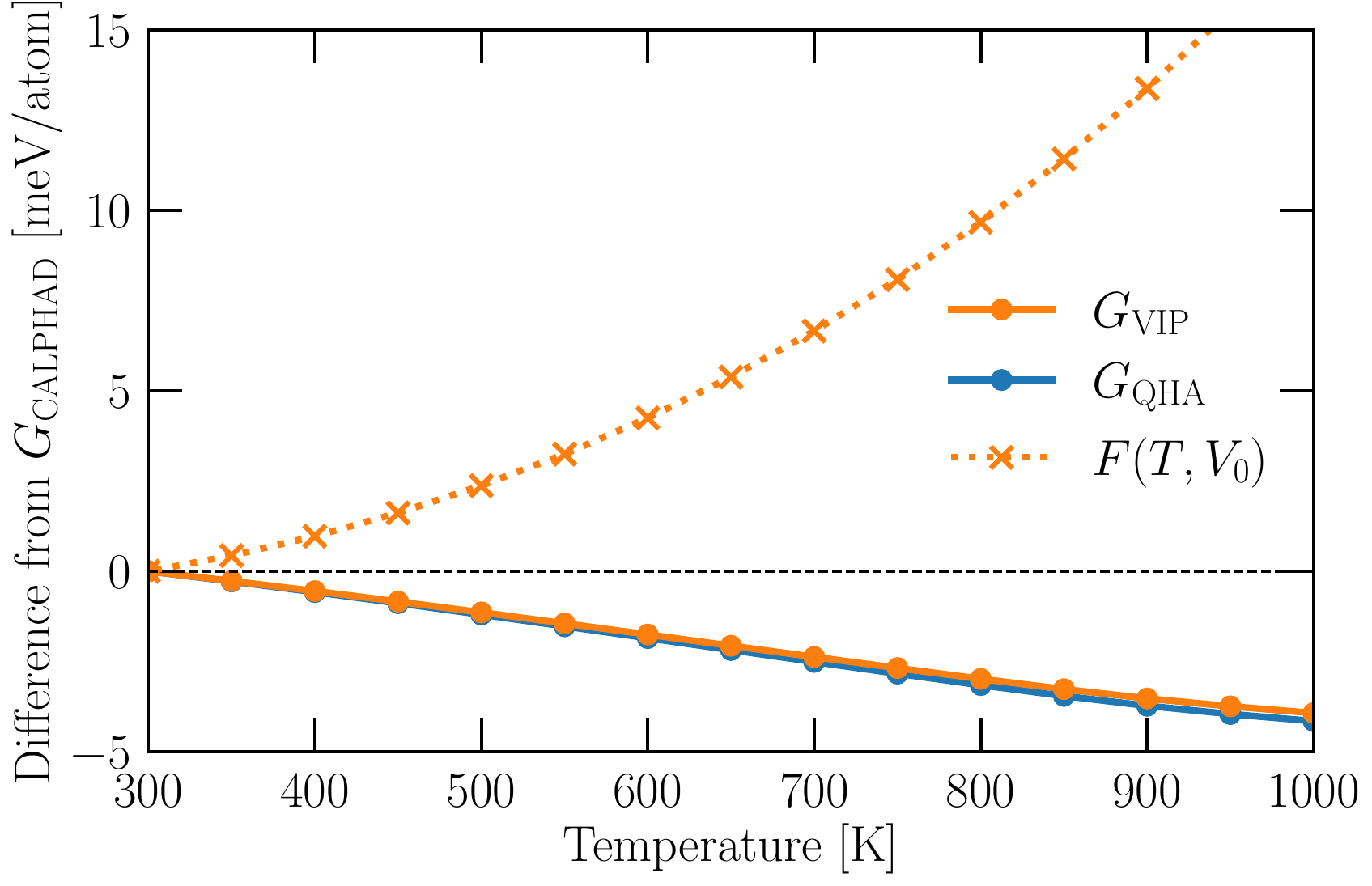}
\caption{\label{fig:Al_calphad_comparison}
Difference in the free energy between the DFT calculations and the CALPHAD data \cite{Dinsdale1991-jk} of fcc Al.
The origin of the DFT free energies is shifted to the CALPHAD data at 300 K.
The orange solid line shows the difference in the Gibbs free energy between the VIP method using the $B_1$, as defined in Eq.~(\ref{B_1}), and the CALPHAD data.
The orange dotted line shows the difference between the Helmholtz free energy $F(T,V_0)$, without thermal expansion, and the Gibbs free energy using the CALPHAD data.
The blue line shows the difference in the Gibbs free energy between the QHA and the CALPHAD data.}
\end{center}
\end{figure}

Differences between the Gibbs energies by DFT and the CALPHAD data are presented in Fig.~\ref{fig:Al_calphad_comparison}.
To examine the temperature dependence, we shifted the origin of the DFT calculations to match the CALPHAD data at 300 K.
Figure~\ref{fig:Al_calphad_comparison} clearly shows that the difference between the Helmholtz free energy $F(T,V_0)$ and the CALPHAD data grows rapidly with increasing temperature, while the Gibbs free energy $G_{\rm VIP}$ exhibits much better agreement due to inclusion of thermal expansion.
This result demonstrates that considering the change in free energy due to thermal expansion is crucial for accurately evaluating phase stability.

\newpage
\subsection{bcc and hcp Ti}
\begin{figure}[b]
  \begin{center}
\includegraphics[width=1.0\linewidth]{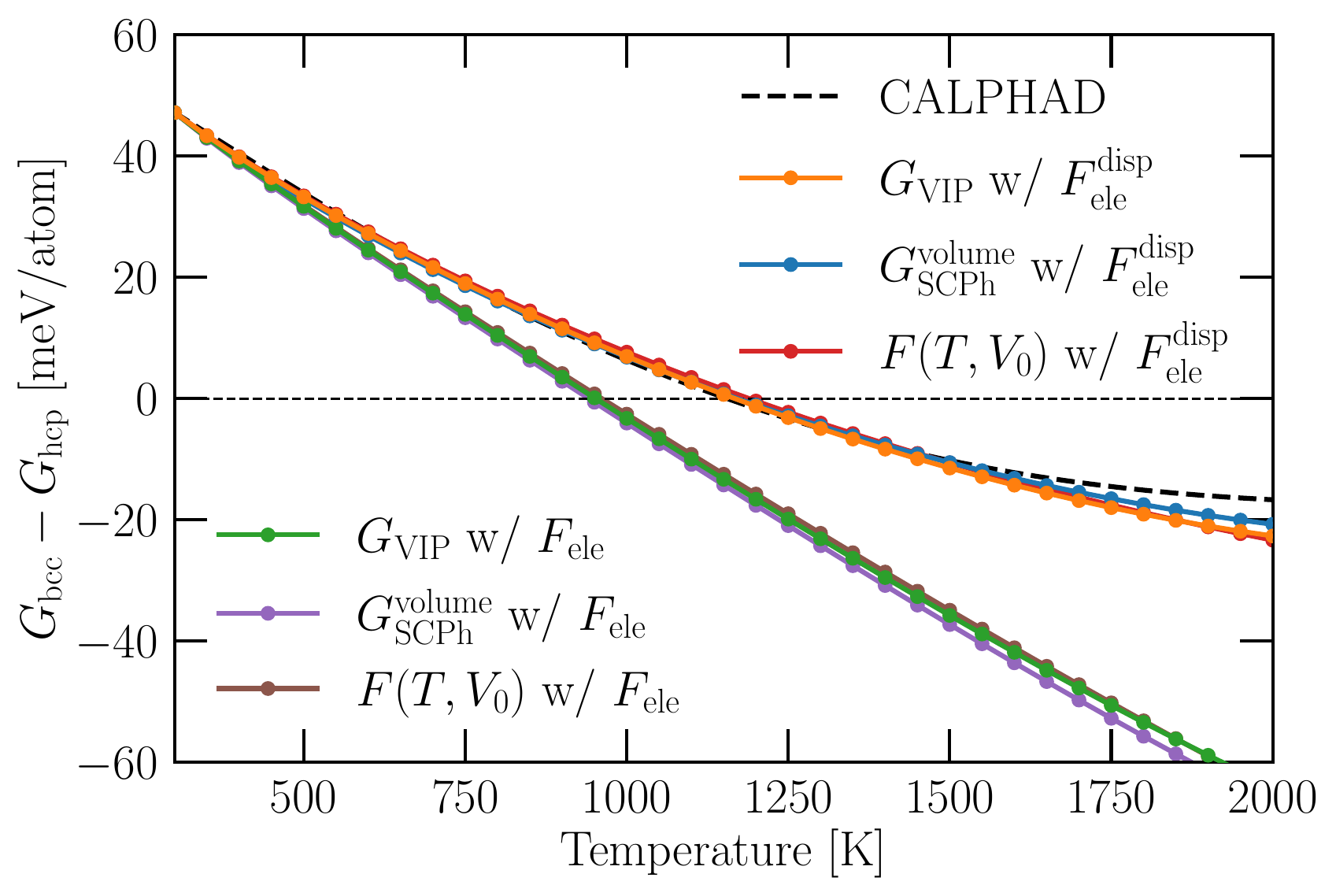}
\caption{
  \label{fig:Ti_Tc}
Difference in the Gibbs free energy between bcc and hcp Ti from 300 to 2000 K.
The origin of the DFT calculations is shifted to the CALPHAD data at 300 K.
At the phase transition temperature $T_{\rm c}$, the vertical axis becomes zero.
The black dashed line represents the Gibbs free energy of the CALPHAD data.
The orange and blue lines represent the differences in the Gibbs free energies between bcc and hcp Ti calculated by using the VIP method and the volume-varied SCPh calculations as denoted by $G_{\rm VIP}$ and $G_{\rm SCPh}^{\rm volume}$, respectively, with vibrational atomic displacements, while the red line represents the difference in the Helmholtz free energy.
The green, purple, and brown lines are same as the orange, blue, and red lines but without vibrational atomic displacements, respectively.
}
\end{center}
\end{figure}

Similar to the discussion on fcc Al, the Gibbs free energy can be accurately calculated without including $B_2$ (Sec. S2 in the Supplemental Material \cite{Supplemental}).
Therefore, in the following calculations, we  used $B_1$ to determine the Gibbs free energy with the VIP method.
Figure~\ref{fig:Ti_Tc} shows the difference in the Gibbs free energy of Ti between bcc and hcp from 300 to 2000 K.
The origin of the vertical axis is shifted that the values from first principles and those of CALPHAD coincides at 300 K, because the GGA overestimates the energy of the bcc phase using the static bcc lattice as a local maximum of the potential energy surface~\cite{Jung2023-yz}.
Both the Gibbs free energy and the Helmholtz free energy follow a similar curve until the phase transition temperature.
Thermal expansion significantly affects the absolute values of the Gibbs free energy as shown in Sec. S4 of the Supplemental Material \cite{Supplemental}.
However, because the thermal expansion contributions to the Gibbs free energy in the bcc and hcp phases are very similar, their effects nearly cancel out when calculating the free energy difference.
In contrast, considering vibrational atomic displacements significantly affects the evaluation of the relative phase stability; the Gibbs free energy leads to a better agreement with the CALPHAD data by including electron-phonon couplings $F_{\rm ele}^{\rm disp}(T,V)$.
Table~\ref{Tc_Ti_list} lists the phase transition temperatures $T_{\rm c}$ of Ti in the different definitions of free energy.
\renewcommand{\arraystretch}{2}
\begin{table}[H]
  \caption{
  \label{Tc_Ti_list}
  Phase transition temperatures $T_{\rm c}$ of Ti in the different definitions of free energy.
  The CALPHAD data is taken from Ref.~\cite{Dinsdale1991-jk}.
  }
  \centering
  \begin{ruledtabular}
    \begin{tabular}{ccccc}
      & \multicolumn{2}{c}{Computational conditions} & \multicolumn{1}{c}{$T_{\rm c}$} & \\
      \hline
      & \multicolumn{2}{l}{$G_{\rm VIP}$ w/ $F_{\rm ele}^{\rm disp}$} & \multicolumn{1}{c}{1167 K} & \\
      & \multicolumn{2}{l}{$G_{\rm SCPh}^{\rm volume}$ w/ $F_{\rm ele}^{\rm disp}$} & \multicolumn{1}{c}{1171 K} & \\ 
      & \multicolumn{2}{l}{$F(T,V_0)$ w/ $F_{\rm ele}^{\rm disp}$} & \multicolumn{1}{c}{1189 K} & \\
      & \multicolumn{2}{l}{$G_{\rm VIP}$ w/ $F_{\rm ele}$} & \multicolumn{1}{c}{952 K} & \\
      & \multicolumn{2}{l}{$G_{\rm SCPh}^{\rm volume}$ w/ $F_{\rm ele}$} & \multicolumn{1}{c}{941 K} & \\
      & \multicolumn{2}{l}{$F(T,V_0)$ w/ $F_{\rm ele}$} & \multicolumn{1}{c}{961 K} & \\
      \hline
      & \multicolumn{2}{l}{CALPHAD} & \multicolumn{1}{l}{1155 K} &
    \end{tabular}
  \end{ruledtabular}
\end{table}
\begin{figure}[b]
  \begin{center}
  \includegraphics[width=1.0\linewidth]{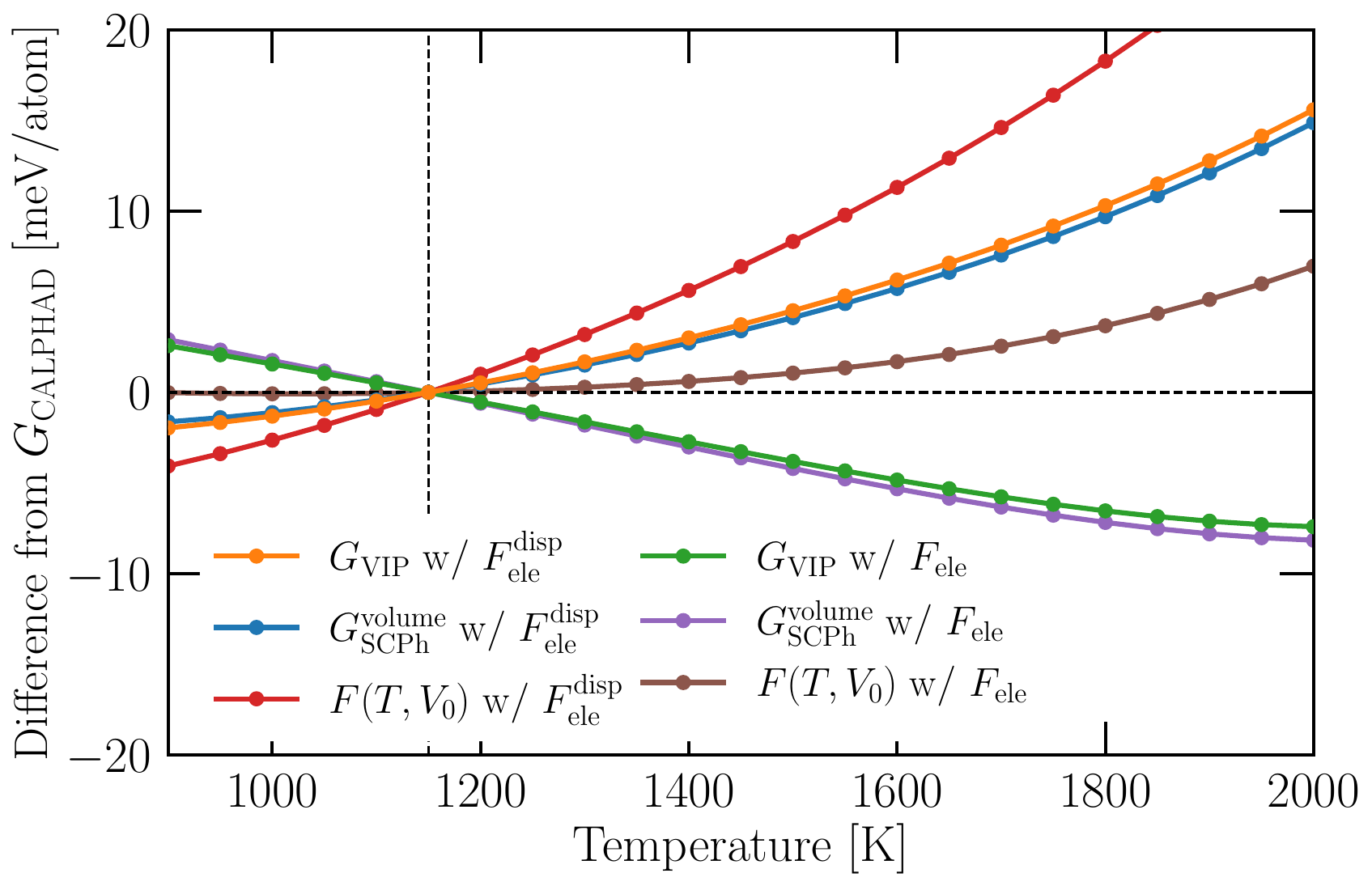}
  \caption{
  \label{fig:bcc_Ti_CALPHAD}
  Difference in the free energy between the DFT calculations and the CALPHAD data \cite{Dinsdale1991-jk} of bcc Ti.
  The origin of the vertical axis is shifted to be zero  at 1150 K that is indicated as the vertical dashed line.
  The orange and blue lines represent the differences in the Gibbs free energies between bcc and hcp Ti calculated by using the VIP method and the volume-varied SCPh calculations as denoted by $G_{\rm VIP}$ and $G_{\rm SCPh}^{\rm volume}$, respectively, with vibrational atomic displacements, while the red line represents the difference in the Helmholtz free energy.
  The green, purple, and brown lines are same as the orange, blue, and red lines but without vibrational atomic displacements, respectively.
  }
  \end{center}
  \end{figure}

Figure~\ref{fig:bcc_Ti_CALPHAD} shows the difference in the free energy between the VIP method and the CALPHAD data \cite{Dinsdale1991-jk} of bcc Ti from 900 to 2000 K.
The  DFT values are shifted to match the CALPHAD data at 1150 K that is indicated by the vertical dashed line.
As observed with Al, incorporating the change in the free energy due to thermal expansion improves the agreement with the CALPHAD data for $G_{\rm VIP}$ including $F_{\rm ele}^{\rm disp}(T,V)$.
Even though the Helmholtz free energy $F(T,V_0)$, which excludes the effect of $F_{\rm ele}^{\rm disp}(T,V)$ appears to reproduce the CALPHAD data better, this agreement can be attributed to merely error cancellation, as  thermal expansion and electron-phonon couplings have opposite tendencies with each other as is clear from the figure.
Namely, electron-phonon couplings increase the Gibbs free energy of the bcc phase.
Particularly, the VIP method accurately reproduces the Gibbs free energies obtained by volume-varied SCPh calculations.

\begin{figure}[t]
\begin{center}
\includegraphics[width=1.0\linewidth]{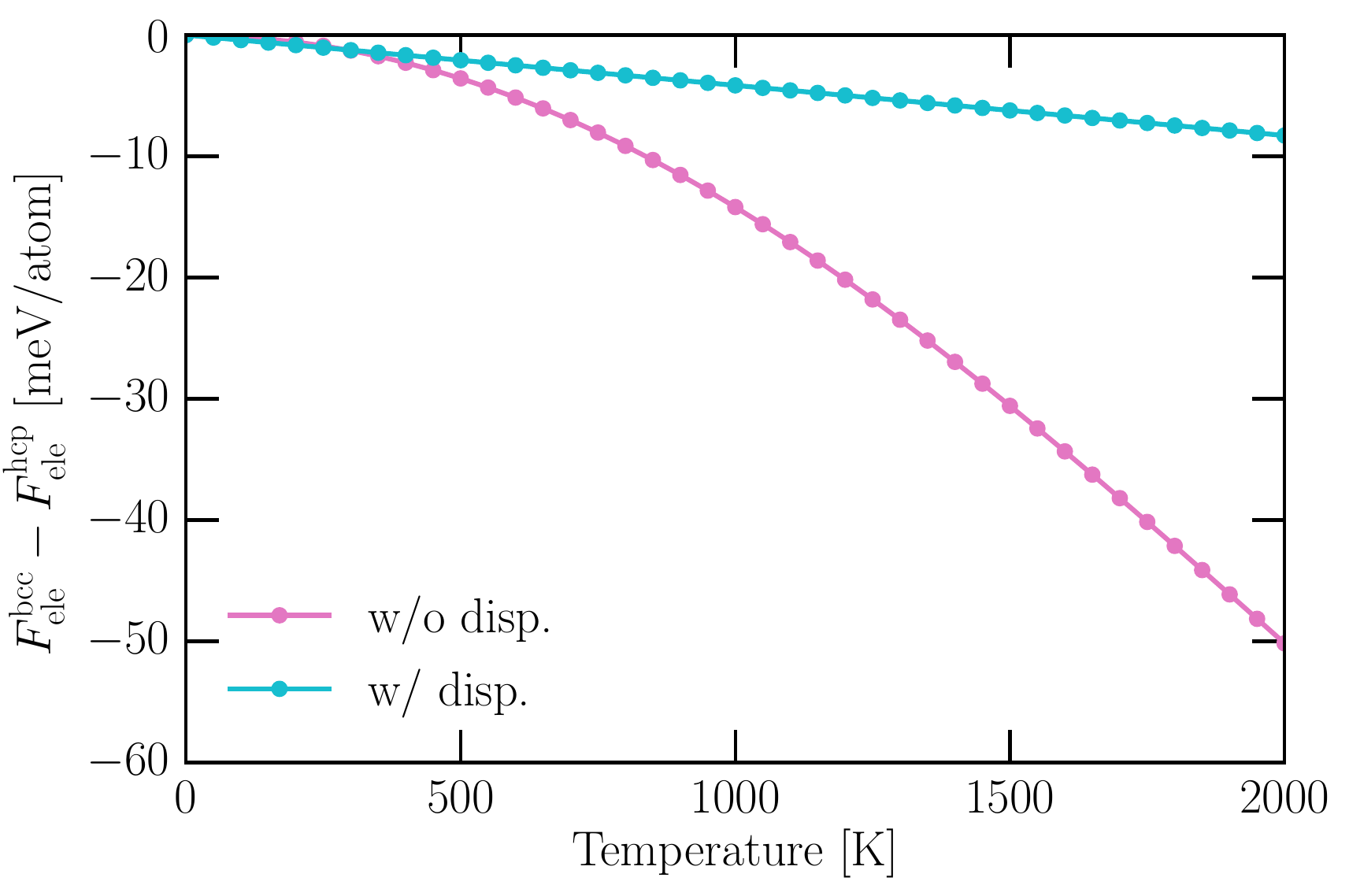}
\caption{
\label{fig:Ti_electronic_free_energy}
Difference in the electronic free energy between bcc and hcp Ti from 0 to 2000 K.
The pink and cyan lines show the difference in the electronic free energy between bcc and hcp Ti, without or with electron phonon couplings, respectively.
The electronic free energies taking $F_{\rm ele}^{\rm disp}(T,V)$ into account by five samplings of vibrational atomic displacements are plotted using the least squares method.
}
\end{center}
\end{figure}
The difference in the electronic free energy between bcc and hcp Ti is shown in Fig.~\ref{fig:Ti_electronic_free_energy}.
As is seen from the figure, the effect of electron-phonon couplings destabilizes the bcc phase, where the difference in the electronic free energy between bcc and hcp Ti decreases from approximately 50 meV/atom to 10 meV/atom.
It should be noted that the maximum errors in the electronic free energy for five vibrational atomic displacements were $\pm$ 0.5 meV/atom for bcc and $\pm$ 0.2 meV/atom for hcp, indicating that even a single sampling may be sufficient.
\begin{figure}[t]
\begin{center}
\includegraphics[width=1.0\linewidth]{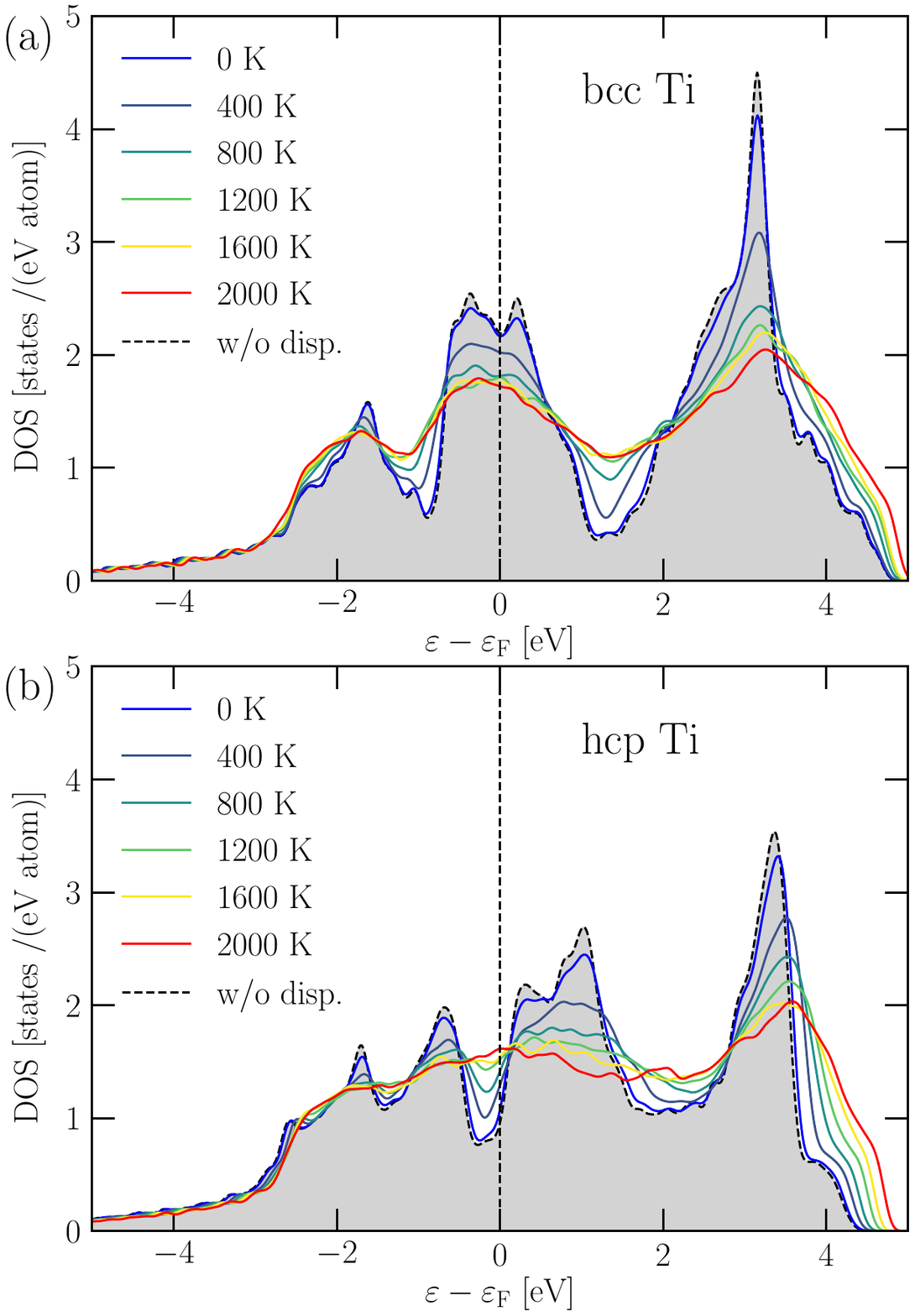}
\caption{
\label{fig:Ti_dos}
Electronic density of states (DOS) of Ti with the (a) bcc and (b) hcp structures.
The top and bottom figures show the DOS of the bcc and hcp phases, respectively.
The blue to red lines show the DOS between 0 and 2000 K.
The blue line represents the DOS at 0 K, which includes zero-point vibration.
The shaded area shows the DOS without vibrational atomic displacements.
}
\end{center}
\end{figure}

In connection with the above discussion, Figs.~\ref{fig:Ti_dos}(a) and (b) show the electronic DOS of bcc and hcp Ti, respectively.
In bcc, the DOS at the Fermi level decreases with increasing temperature.
In contrast, the DOS at the Fermi level increases with increasing temperature in hcp.
A particularly notable point is that the randomization of atomic coordinates leads to a smoother DOS with increasing temperature.
Although the 3$d$ valence electrons of Ti are localized, the randomization of atomic coordinates due to vibrational atomic displacement leads to a smoother DOS and significant changes in the DOS at the Fermi level.
The electronic free energy difference between the bcc and hcp phases, caused by changes in the DOS due to vibrational atomic displacements, i.e., electron-phonon couplings, plays a pivotal role in determining the phase transition temperature.
Omitting this contribution leads to substantial discrepancies, particularly at elevated temperatures.
Finally, we present the computational costs for calculating the Gibbs free energy of bcc and hcp Ti within SCPh theory using the VIP method, as detailed in Table~\ref{CPUtimes}, which indicates a significant advantage of our method.
\begin{table}[H]
\caption{
\label{CPUtimes}
Estimated CPU times (in core-hours) for bcc Ti, hcp Ti, and tetragonal ZrO$_2$ of DFT calculations.
The SCPh calculations were implemented for seven volumes.
The CPU times were obtained using an AMD EPYC 7702 2.0 GHz 64-core processor in ISSP supercomputers.
}
\centering
\begin{ruledtabular}
{\renewcommand{\arraystretch}{2}
\begin{tabular}{ccc}
\multicolumn{3}{c}{CPU time (core hour) of DFT calculations}\\
\hline
& VIP & volume-varied SCPh\\
\cline{1-3}
bcc Ti & 243 & 1,953\\
hcp Ti & 348 & 1,860\\
tetragonal ZrO$_2$ & 186 & 1,432\\
\end{tabular}
}
\end{ruledtabular}
\end{table}

\subsection{tetragonal ZrO\texorpdfstring{$_2$}{2}}
\begin{figure}[H]
  \begin{center}
  \includegraphics[width=1.0\linewidth]{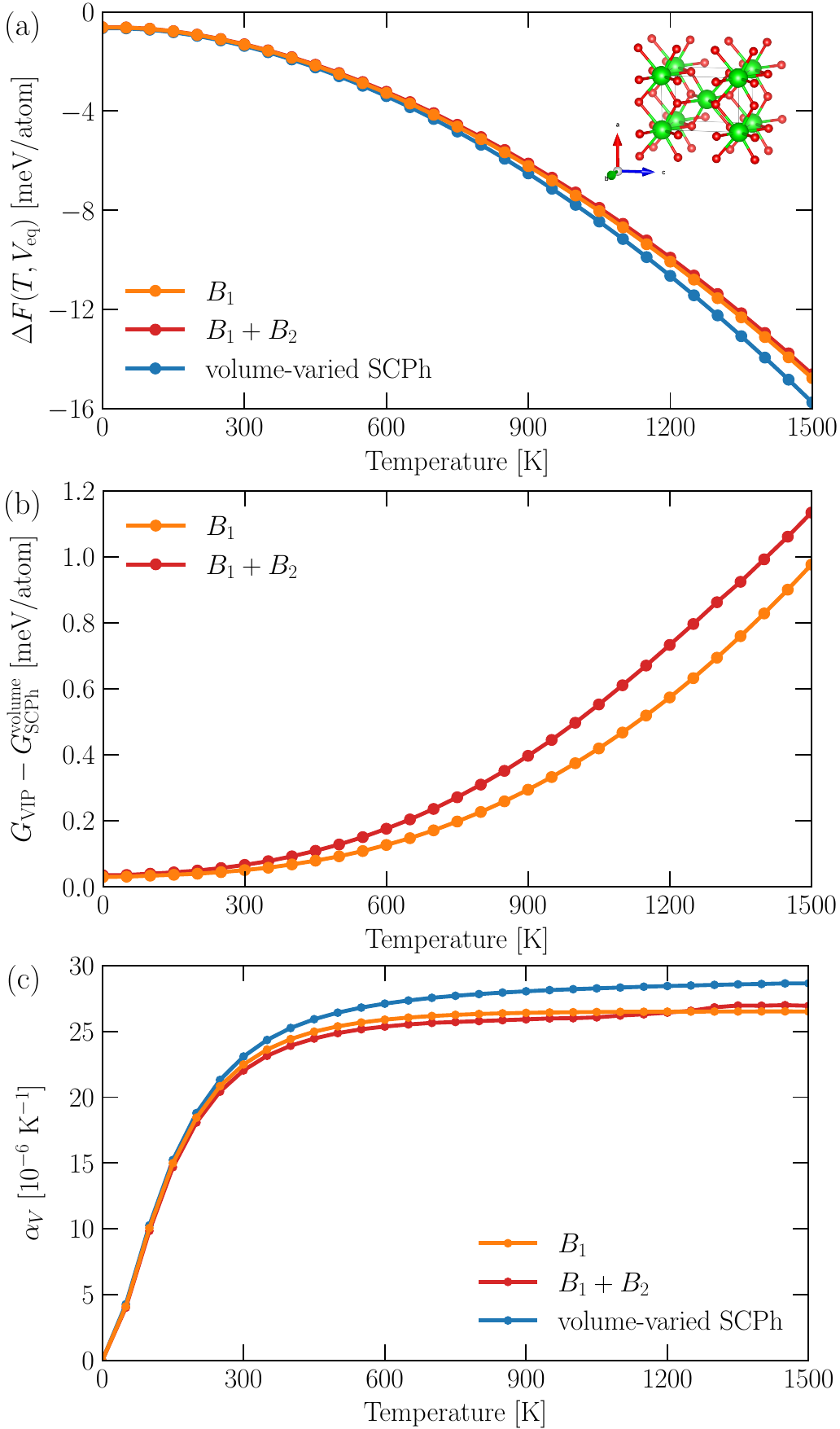}
  \caption{
  \label{fig:t-ZrO2}
  Comparison between the VIP method and the volume-varied SCPh calculations in terms of the free energy and the CVTE for tetragonal ZrO$_2$ over the temperature range from 0 to 1500 K.
  (a) The change in the Helmholtz free energy due to thermal expansion $\Delta F=-\int_{V_0}^{V_{\rm eq}} P(T,V)dV$ in Eq.~(\ref{gibbs_trans}).
  (b) The difference in the Gibbs free energy between the VIP method and the volume-varied SCPh calculations $G_{\rm SCPh}^{\rm volume}$.
  (c) The CVTE $\alpha_V = 1/ V_{\rm eq} \left({dV_{\rm eq}}/{dT}\right)$ where $V_{\rm eq}$ is the equilibrium volume at each temperature.
  The orange line is the standard VIP result ($B_1$ only, Eq.~(\ref{B_1})); the red line is a reference that also includes $B_2$ ($B_1+B_2$, Eqs.~(\ref{B_1}) and (\ref{B_2})).
  The blue lines represent the CVTE obtained by volume-varied SCPh calculations.
  The maximum deviation of the CVTE between the VIP method and the volume-varied SCPh calculations is about 7.5\% at 1500 K.
  The crystal structure shown as inset of (a) was generated using VESTA \cite{Momma2011-zl}.
  }
\end{center}
\end{figure}
Figure~\ref{fig:t-ZrO2} shows the comparison between the VIP method and the volume-varied SCPh calculations for tetragonal ZrO$_2$ in terms of the free energy and the coefficient of volumetric thermal expansion (CVTE) over the temperature range from 0 to 1500 K.
The difference in the Gibbs free energy between the VIP method and the volume-varied SCPh calculations is less than 1 meV/atom over the temperature range from 0 to 1500 K.
This figure demonstrates the applicability of the VIP method to accurately and efficiently reproduce the results obtained by volume-varied SCPh calculations even for tetragonal ZrO$_2$ with six atoms in the primitive cell.
Calculated CVTE exhibits the deviation of approximately 7.5\% from the volume-varied SCPh at 1500 K, but improved accuracy is also observed at lower temperatures (e.g., deviation approximately 4.5\% at 600 K).
We employed $c/a$-varied SCPh calculations that independently varied the $a$ and $c$ axes to minimize the Gibbs free energy at each temperature.
The resulting $c/a$ ratio can be considered constant in the temperature range of 0 to 1500 K, with a maximum change of less than 0.06\% from 0 K.
The results of the $c/a$-varied SCPh calculations for tetragonal ZrO$_2$ in terms of the CLTE and the CVTE over the temperature range from 0 to 1500 K are shown in Sec. S5 of the Supplemental Material \cite{Supplemental}.
Table~\ref{CPUtimes} also lists the computational costs for calculating the Gibbs free energy of tetragonal ZrO$_2$ within SCPh theory using the VIP method.

\section{Summary}
We have proposed a method, i.e., the VIP method, for calculating the Gibbs free energy from constant-volume phonon calculations.
A notable advantage of our method compared with the conventional QHA is its ability to calculate vibrational pressure and its volume derivative from single-volume first-principles phonon calculations using Gr\"{u}neisen parameters resulting in significant reduction in computational costs.
We applied our method to fcc Al exhibiting significant thermal expansion, where the Gibbs free energy and the linear coefficient of thermal expansion calculated by the VIP method agree well with the QHA and experimental results.
Furthermore, we also applied our method to bcc and hcp Ti, where effects of thermal expansion and electron-phonon couplings nearly cancel out in evaluating the Gibbs free energy of the bcc phase.
Considering both effects, the hcp-bcc phase-transition temperature agreed satisfactorily with experimental results.
Moreover, we applied the VIP method to tetragonal ZrO$_2$.
In these cases, the Gibbs free energy and the coefficient of linear thermal expansion calculated by the VIP method agree well with the volume-varied SCPh calculations.
Due to low computational costs of our method, we expect high-throughput first-principles phase diagram calculations will become possible opening up new pathway for materials design considering phase equilibria.
Such an approach should also be reinforced by, e.g., data assimilation techniques.

\begin{acknowledgments}
This work was partly supported by JSPS-KAKENHI Grant No. JP24K01144 and by MEXT-DXMag Grant No. JPMXP1122715503.
The calculations were partly carried out by using supercomputers at ISSP, The University of Tokyo, and TSUBAME4.0, Institute of Science Tokyo.
\end{acknowledgments}


\begingroup
\appendix
\section{FORCE CONSTAMTS AND GR\"{U}NEISEN PARAMETER}
\label{Gruneisen_parameter_definitions}

If the vibrational displacements are small compared with the interatomic distance, the potential energy $U(\{\boldsymbol{R}_i\})$ where $\boldsymbol{R}_i$ is the atomic position of the interacting atomic system can be expanded in a power series of the displacements $\boldsymbol{u}_i=\boldsymbol{R}_i-\boldsymbol{R}_i^0$ as follows:
\begin{align}
U=U_0&+U_2+U_3+U_4+\dots,\\
U_n=\frac{1}{n!}&\sum_{\{l,\kappa,\mu\}}\Phi_{\mu_1,\dots,\mu_n}(l_1\kappa_1;\dots;l_n\kappa_n)\nonumber\\
&\times u_{\mu_1}(l_1\kappa_1){\cdots}u_{\mu_n}(l_n\kappa_n),
\label{pes}
\end{align}
where $\mu=x,y,z$ and $u_{\mu}(l\kappa)$ is the displacement of atom $\kappa$ in the $l$th unit cell.
In. Eq.~(\ref{pes}) the linear term $U_1$ is omitted because atomic forces are zero in equilibrium positions.
The coefficient $\Phi_{\mu_1,\dots,\mu_n}(l_1\kappa_1,\dots,l_n\kappa_n)$ defined as
\begin{align}
\Phi_{\mu_1,\dots,\mu_n}&(l_1\kappa_1;\dots;l_n\kappa_n)\nonumber\\
&=\left.\frac{\partial^nU}{\partial{u}_{\mu_1}(l_1\kappa_1)\cdots\partial{u_{\mu_n}(l_n\kappa_n)}}\right|_{\{\boldsymbol{R}\}=\{\boldsymbol{R}^0\}},
\end{align}
are called the $n$th order interatomic force constants in real-space representation.
The phonon frequency $\omega_{\boldsymbol{q}j}$ can be obtained by diagonalizing dynamical matrix $D$ as
\begin{align}
&\sum_{\kappa'\mu'}D(\kappa\kappa';\boldsymbol{q})\boldsymbol{e}_{\mu'}(\kappa';\boldsymbol{q}j)=\omega_{\boldsymbol{q}j}^2\boldsymbol{e}_{\mu}(\kappa;\boldsymbol{q}j),\\
&{D}_{\mu\nu}(\kappa\kappa';\boldsymbol{q})=\frac{1}{\sqrt{M_{\kappa}M_{\kappa'}}}\sum_{l'}{\Phi_{\mu\nu}}(0\kappa;l'\kappa')e^{i\boldsymbol{q}\cdot\boldsymbol{r}(l')},
\end{align}
where $\boldsymbol{e}(\kappa;\boldsymbol{q}j)$ is the polarization vector and $M_{\kappa}$ is the atomic mass of atom $\kappa$ and $\boldsymbol{r}(l)$ is the translational vector of the $l$-th unit cell. 
Gr\"{u}neisen parameter in Eq.~(\ref{gruneisen}) is derived from the change in the dynamical matrix $\delta{D}$ \cite{Ashcroft2011,Srivastava2023-uy} as
\begin{align}
\gamma_{\boldsymbol{q}j}(V)
&=-\frac{(\boldsymbol{e}_{\boldsymbol{q}j}^*)^{\rm{T}}\delta{D}(\boldsymbol{q})\boldsymbol{e}_{\boldsymbol{q}j}}{6\omega_{\boldsymbol{q}j}^2(V)},\\
\delta{D}_{\mu\nu}(\kappa\kappa';\boldsymbol{q})
&=\frac{1}{\sqrt{M_{\kappa}M_{\kappa'}}}\sum_{l'}\delta{\Phi_{\mu\nu}}(l\kappa;l'\kappa')e^{i\boldsymbol{q}\cdot\left(\boldsymbol{r}(l')-\boldsymbol{r}(l)\right)},\\
\delta{\Phi}_{\mu\nu}(l\kappa,l'\kappa')
&=\sum_{l'',\kappa'',\lambda}\Phi_{\mu\nu\lambda}(l\kappa;l'\kappa';l''\kappa'')r_{\lambda}(l''\kappa'').
\end{align}

\endgroup

\newpage
\clearpage
\onecolumngrid

\begingroup

\makeatletter
\ifdefined\@chapapp
  \gdef\@chapapp{}
\fi
\makeatother

\renewcommand{\thesection}{S\arabic{section}}
\renewcommand{\theequation}{S\arabic{equation}}
\renewcommand{\thefigure}{S\arabic{figure}}
\renewcommand{\thetable}{S\arabic{table}}
\setcounter{section}{0}
\setcounter{equation}{0}
\setcounter{figure}{0}
\setcounter{table}{0}

\section*{Supplemental Material}
\input{Supplemental_Material.tex}

\endgroup

\end{document}

%% file: Supplemental_Material.tex
\title{Efficient first-principles approach to Gibbs free energy with thermal expansion
\\
Supplemental Materials}

\author{Kota Hashimoto}
 \email{hashimoto.k.d594@m.isct.ac.jp}

\author{Tomonori Tanaka}

\author{Yoshihiro Gohda}
 \email{gohda@mct.isct.ac.jp}

\affiliation{
Department of Materials Science and Engineering, Institute of Science Tokyo, Yokohama 226-8501, Japan
}

\onecolumngrid
\section{Temperature dependence of the \texorpdfstring{$c/a$}{c/a} in hcp Ti}
\begin{figure}[H]
  \begin{center}
    \includegraphics[width=0.8\linewidth]{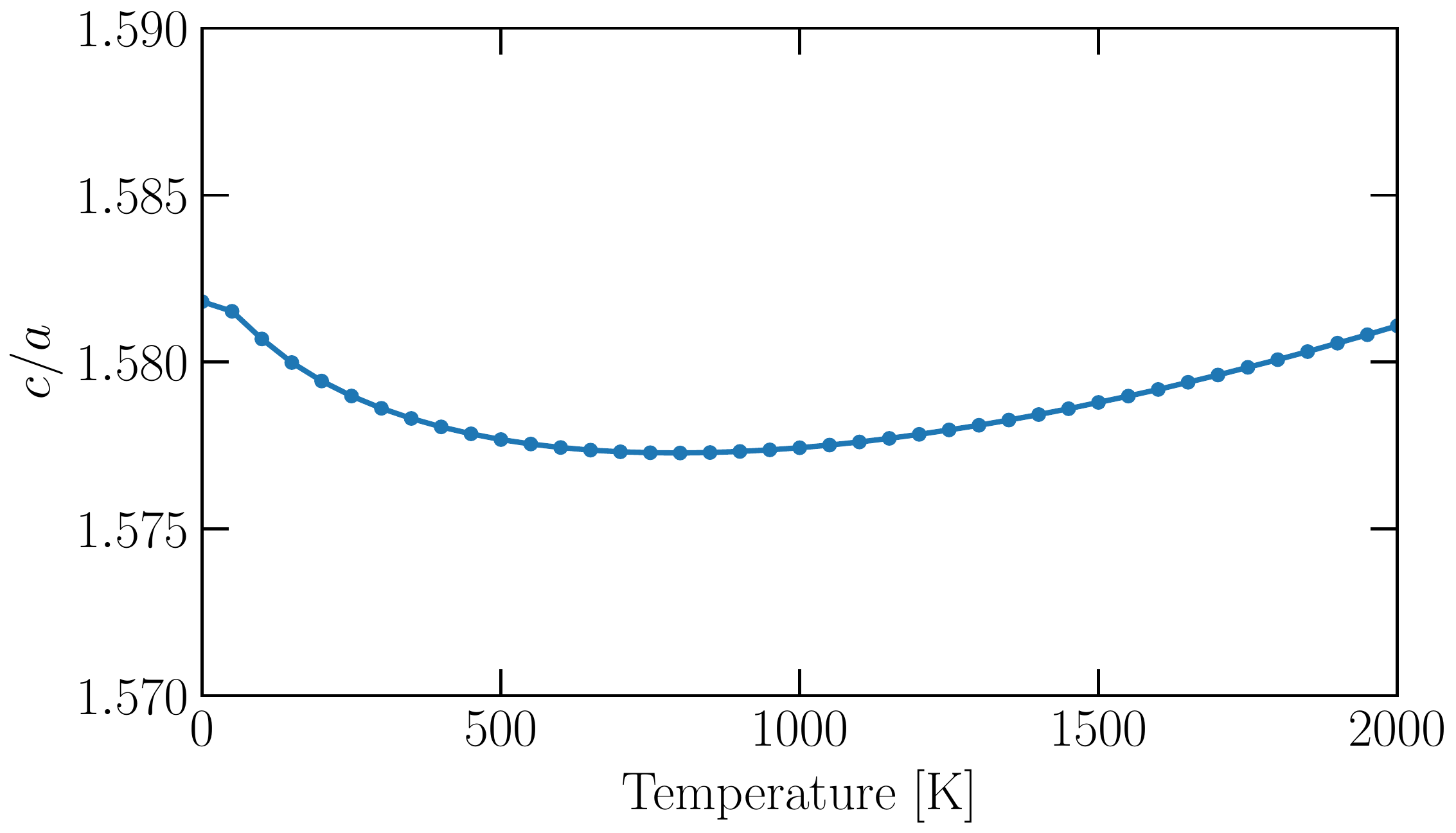}
    \caption{
      \label{fig:c_a}
      Temperature dependence of the $c/a$ ratio in hcp Ti calculated using the QHA.
      Phonon calculations were performed using the harmonic approximation, and the effect of electron-phonon coupling was not considered.
      }
    \end{center}
\end{figure}

Figure~\ref{fig:c_a} illustrates the temperature dependence of the $c/a$ ratio in hcp Ti.
The $c/a$ ratio was calculated using the QHA.
The maximum change in the $c/a$ ratio from 0 K is about 0.3 \%; therefore, approximating the $c/a$ ratio as a constant is reasonable.
$\\$
$\\$

\newpage
\section{Contributions of \texorpdfstring{$B_2$}{B2} for Gibbs free energy}
\label{B_2_term}
\begin{figure}[H]
  \begin{center}
    \includegraphics[width=1.0\linewidth]{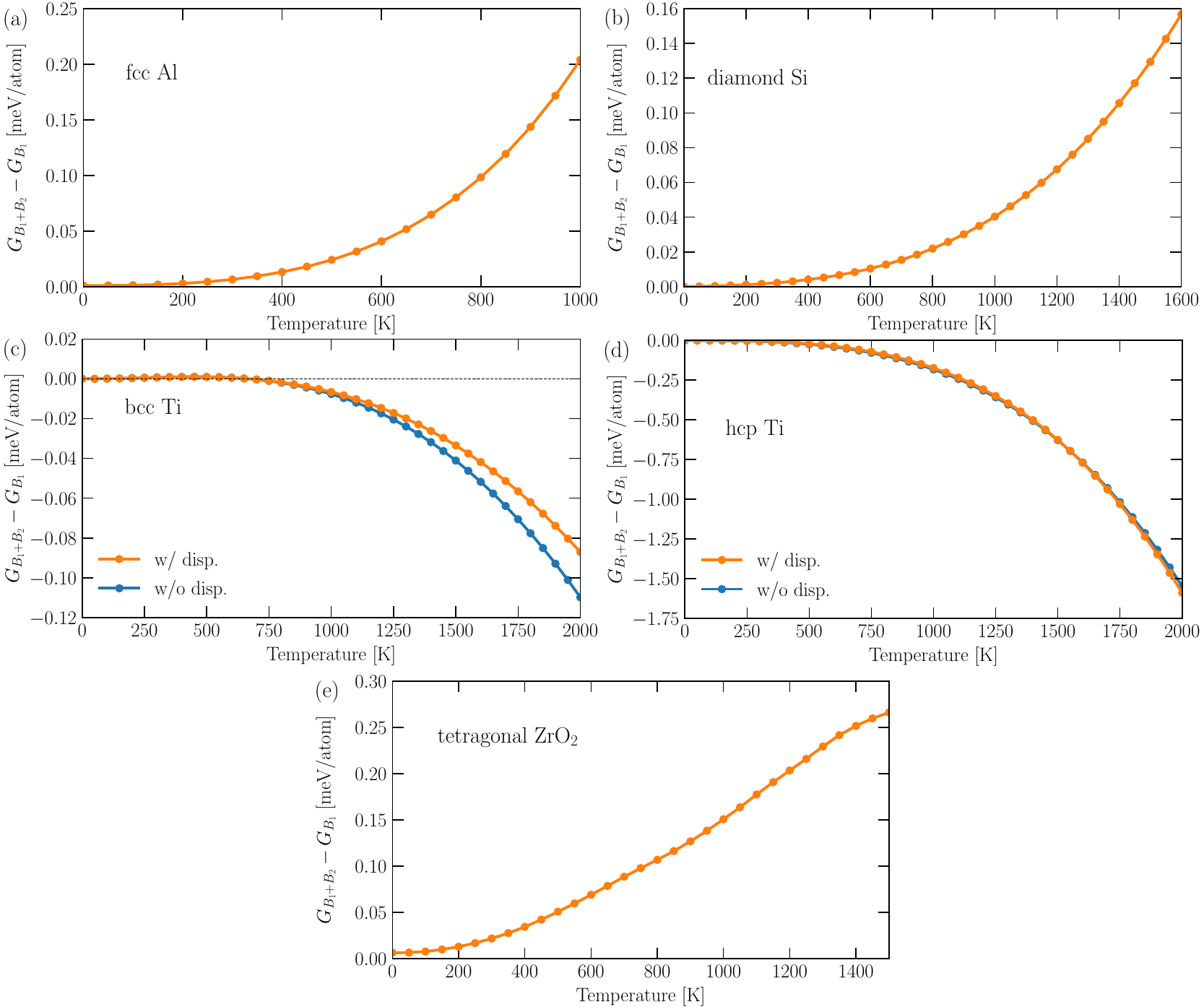}
    \caption{
      \label{fig:B2_expansion}
      Difference in the Gibbs free energy between $B_1$ and $B_1+B_2$, calculated using the VIP method, for (a) fcc Al, (b) diamond Si, (c) bcc Ti, (d) hcp Ti, and (e) tetragonal ZrO$_2$, respectively.
      The $B_2$ term was calculated by explicitly varying the volume for benchmarking purposes.
      In bcc Ti and hcp Ti, the orange and blue lines represent, respectively, the differences in the Gibbs free energy obtained when the effect of vibrational atomic displacements on the electronic density of states is considered and when it is not considered.
      }
    \end{center}
\end{figure}

Figure~\ref{fig:B2_expansion} shows the differences in the Gibbs free energy between $B_1$ and $B_1+B_2$, calculated using the VIP method, for fcc Al, diamond Si, bcc Ti, hcp Ti, and tetragonal ZrO$_2$, respectively.
These results show that the contribution of $B_2$ is negligible in all cases.
$\\$
$\\$

\newpage
\onecolumngrid
\section{Convergence of the second-order Birch-Murnaghan equations of state for fcc Al}
\begin{figure}[H]
  \begin{center}
  \includegraphics[width=0.5\linewidth]{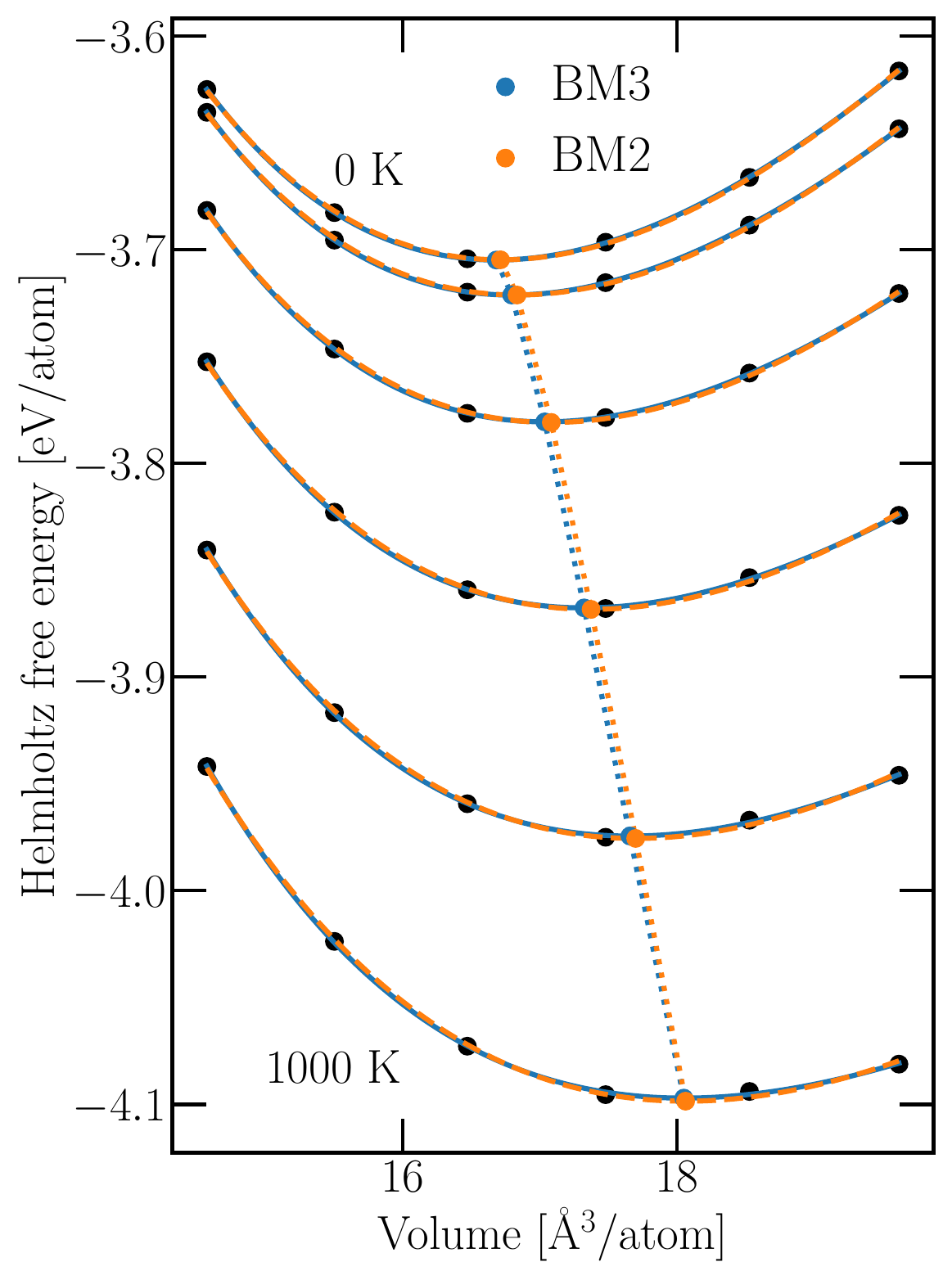}
  \caption{
  \label{fig:eos_difference_Al}
  Difference between the fitting curves obtained using the second- and third-order Birch-Murnaghan equations of state for fcc Al, calculated in 200 K intervals from 0 to 1000 K.
  The blue and orange lines represent the fitted results obtained using the third-order (BM3) and second-order (BM2) Birch-Murnaghan equations of state, respectively.
  }
  \end{center}
\end{figure}

Figure~\ref{fig:eos_difference_Al} shows the difference between the fitting curves obtained using the second- and third-order Birch-Murnaghan equations of state for fcc Al.
The second-order Birch-Murnaghan EOS converges well onto DFT data around 15 \% of the reference volume, which is sufficient for our calculations.
$\\$

\onecolumngrid
\section{The sensitivity of the VIP method to the choice of reference volume}
\begin{figure}[H]
  \begin{center}
  \includegraphics[width=0.8\linewidth]{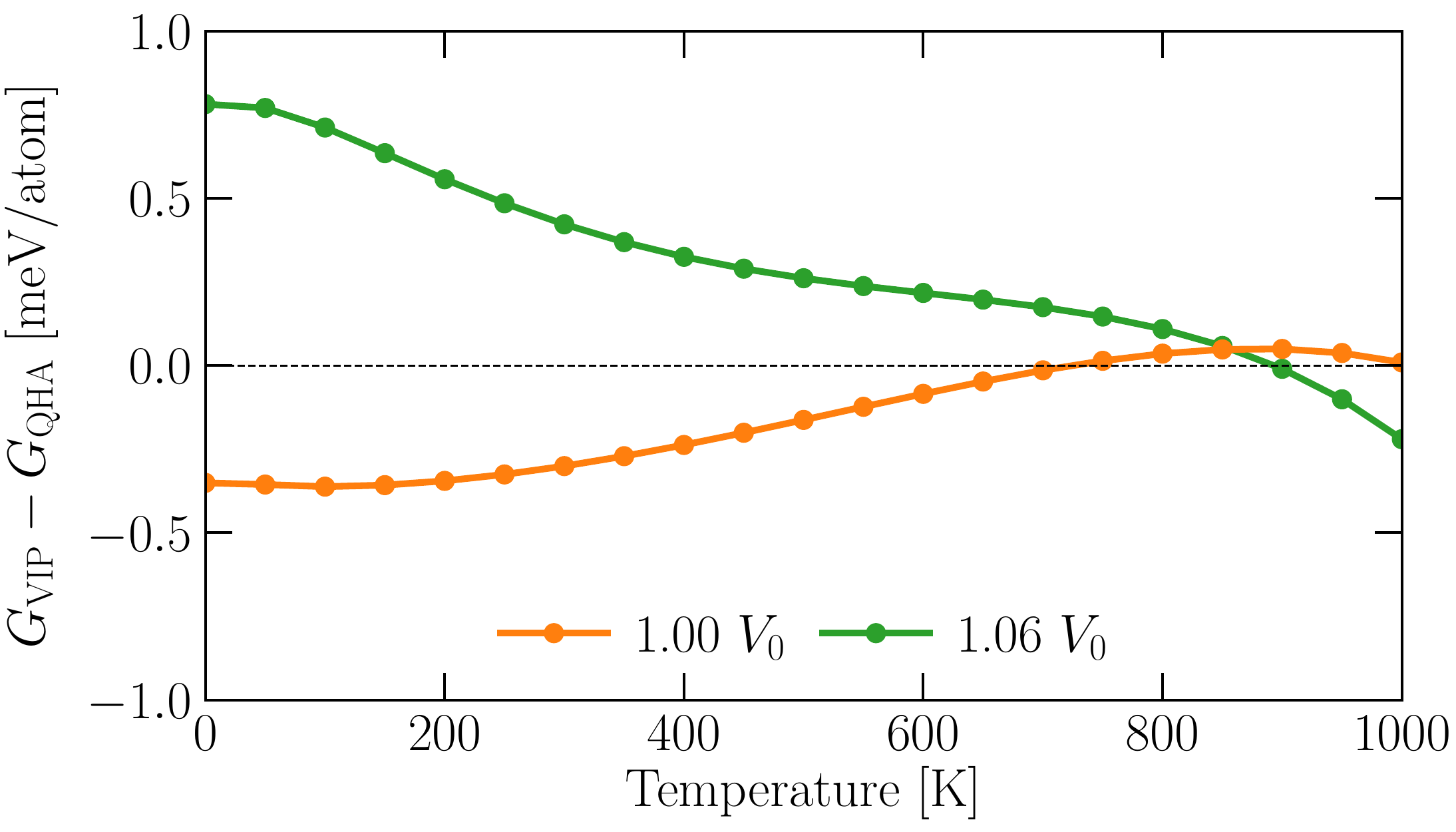}
  \caption{
  \label{fig:initial_volume_fccAl}
  Difference in the Gibbs free energy between the VIP method using $B_1$ and the QHA of fcc Al from 0 to 1000 K for different reference volumes of 1.00 $V_0$ (orange) and 1.06 $V_0$ (green), respectively.
  }
  \end{center}
\end{figure}

\begin{figure}[H]
  \begin{center}
  \includegraphics[width=1.0\linewidth]{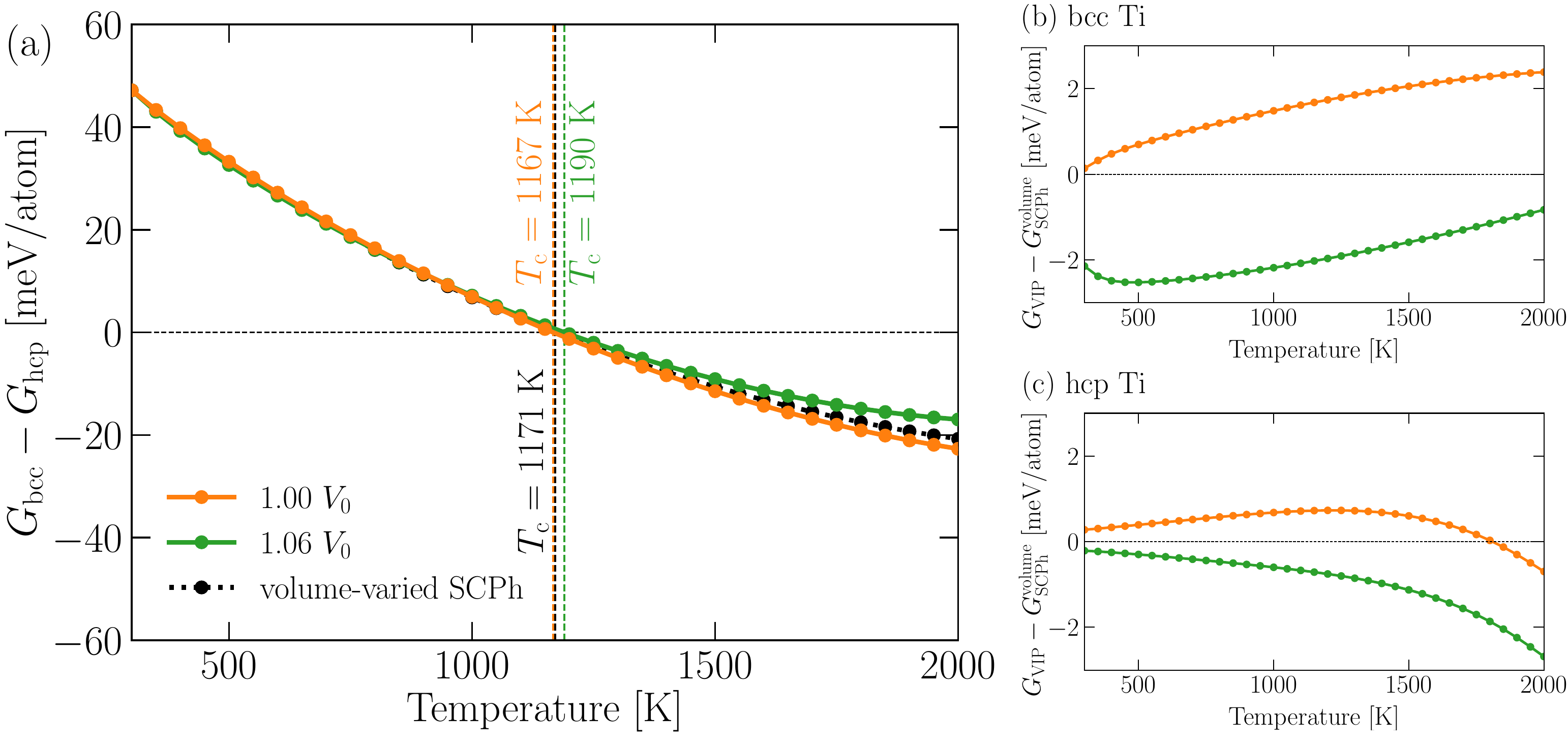}
  \caption{
  \label{fig:initial_volume_bccTi}
  (a) Difference in the Gibbs free energy between bcc and hcp Ti from 300 to 2000 K for the VIP method using $B_1$ with 1.00 $V_0$, 1.06 $V_0$, and the volume-varied SCPh calculations.
  The origin of the DFT calculations is shifted to the CALPHAD data \cite{Dinsdale1991-jk} at 300 K.
  Vibrational atomic displacements were incorporated into the electronic free energy calculations.
  (b) Difference in the Gibbs free energy between the VIP method and the volume-varied SCPh calculations $G_{\rm SCPh}^{\rm volume}$ for bcc Ti.
  (c) same as (b) but for hcp Ti.
  }
  \end{center}
\end{figure}

We have confirmed the sensitivity to the selection of the reference volume for both fcc Al, which exhibits significant thermal expansion, and bcc Ti, which exhibits pronounced anharmonicity.
We used an expanded reference volume of 1.06 $V_0$ (i.e., a 6 \% expansion relative to $V_0$), where $V_0$ was obtained from structural optimization at zero temperature and zero external pressure.
The difference in the Gibbs free energy between the VIP method using $B_1$ and the QHA for fcc Al is shown in Fig.~\ref{fig:initial_volume_fccAl}.
The Gibbs free energy calculated using the VIP method when selecting 1.06 $V_0$ as the reference volume reproduces the QHA results to within 1 meV/atom across the temperature range.
Figure~\ref{fig:initial_volume_bccTi} (a) shows the difference in the Gibbs free energy between bcc and hcp Ti from 300 to 2000 K.
As shown in Fig.~\ref{fig:initial_volume_bccTi} (b) and (c), the use of 1.06 $V_0$ changes the Gibbs free energy for the bcc and hcp phases by approximately 2 meV/atom and 1 meV/atom, respectively, over the temperature range.
Moreover, this selection has a negligible effect on the evaluation of the phase transition temperature.

\newpage
$\\$

$\\$

\clearpage
\onecolumngrid
\section{Applications}
In this section, we present the results obtained by the VIP method for various materials.
We explicitly calculated the $B_2$ term in Eq.~(17) by performing calculations at different volumes for benchmarking purposes in the VIP method.
These figures show that the VIP method can accurately reproduce the results obtained by the QHA and the volume-varied SCPh calculations.
$\\$

\subsection{diamond Si}
\begin{figure}[H]
  \begin{center}
  \includegraphics[width=0.5\linewidth]{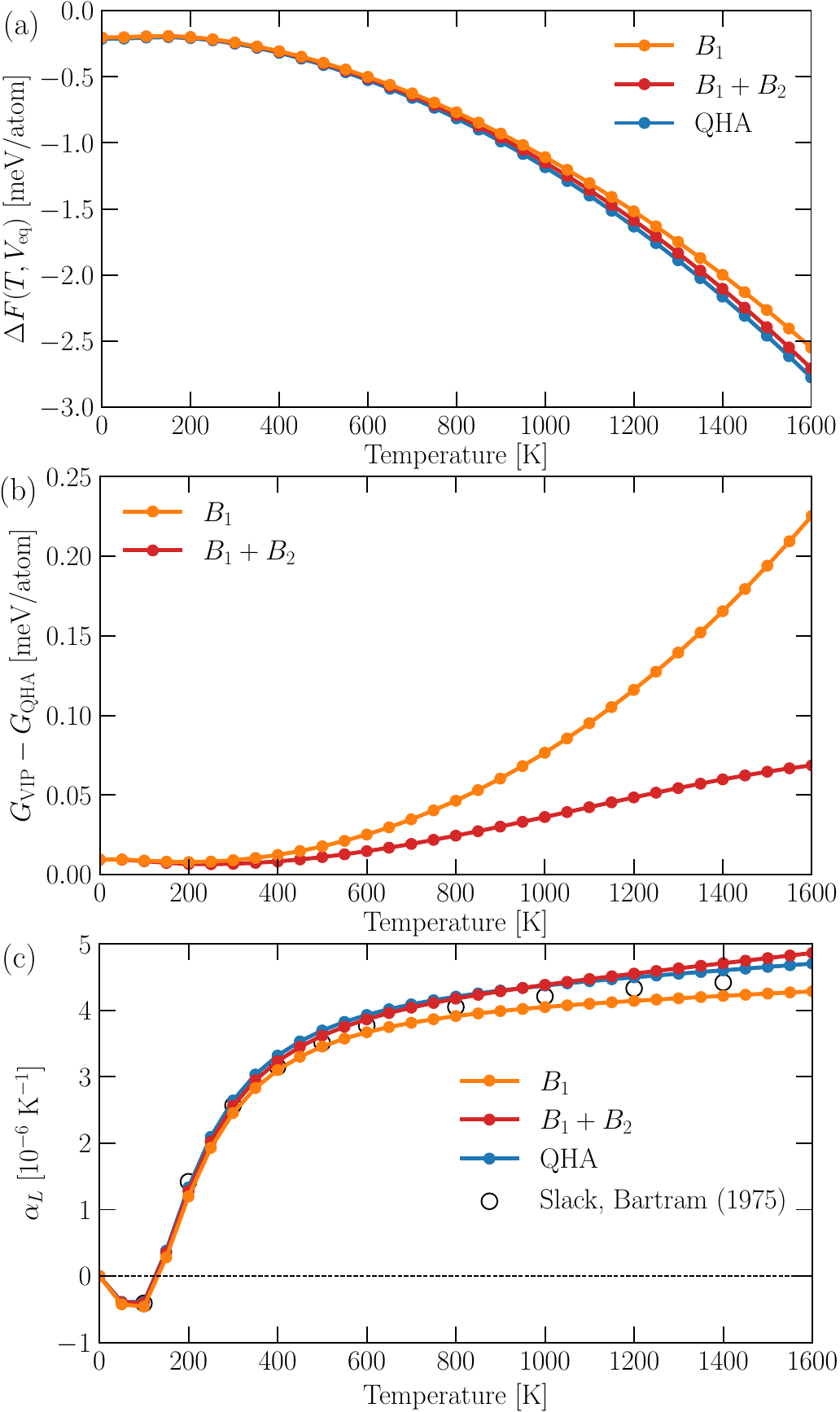}
  \caption{
  \label{fig:Si}
  Comparison between the VIP method and the QHA for diamond Si.
  (a) The change in the Helmholtz free energy due to thermal expansion.
  (b) The difference in the Gibbs free energy between the VIP method and the QHA.
  (c) The CLTE $\alpha_L$.
  The experimental data are taken from Ref.~\cite{Slack1975-dy}.
  The orange line is the standard VIP result ($B_1$ only, Eq.~(16)); the red line is a reference that also includes $B_2$ ($B_1+B_2$, Eqs.~(16) and (17)).
  The blue line represents the result of the QHA.
  }
  \end{center}
\end{figure}

The comparison between the VIP method and the QHA for diamond Si in terms of the free energy and the coefficient of linear thermal expansion (CLTE) are shown in Fig.~\ref{fig:Si}.
$\\$

\clearpage
\onecolumngrid
\subsection{Ti}
\subsubsection{bcc Ti}
\begin{figure}[H]
  \begin{center}
  \includegraphics[width=1.0\linewidth]{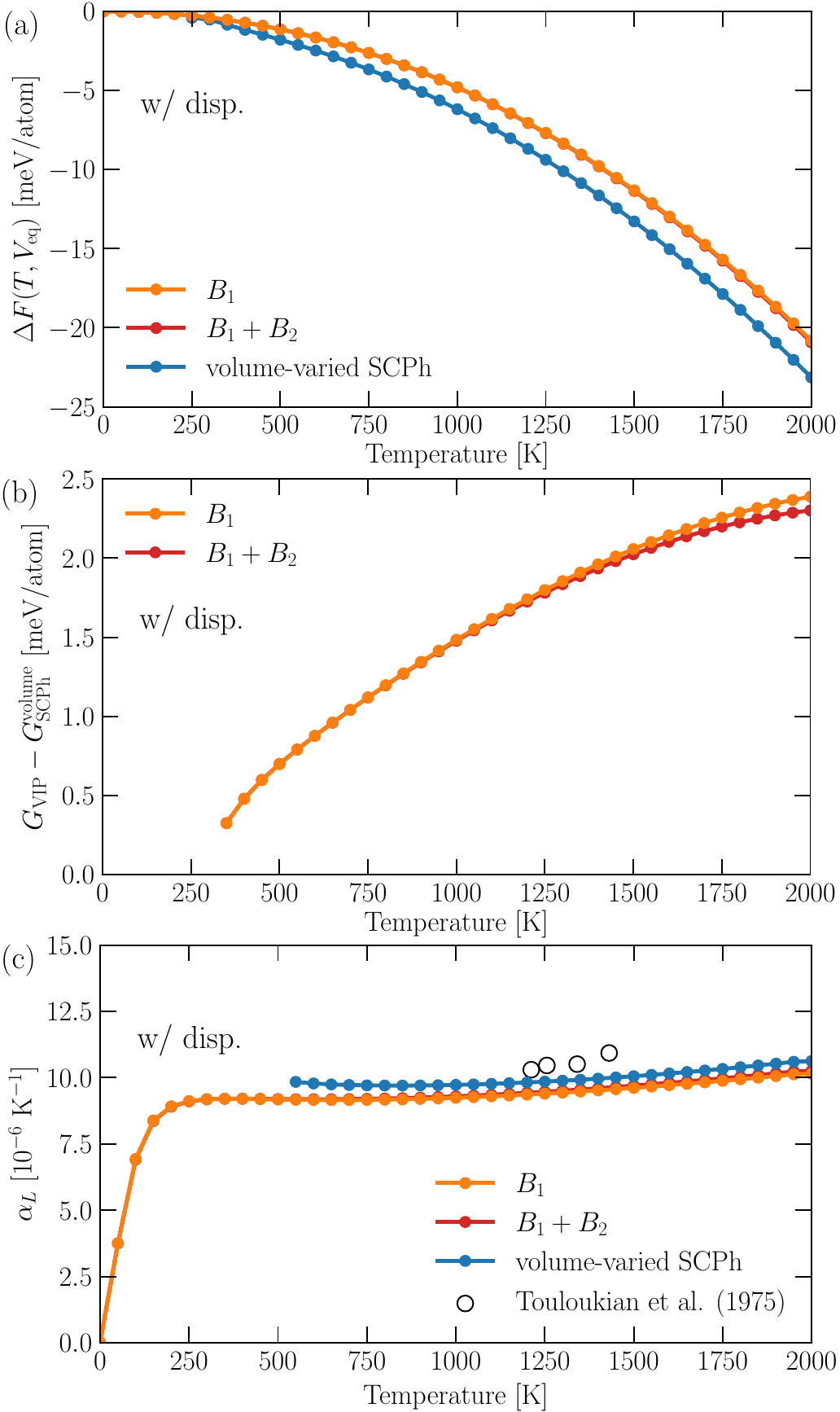}
  \caption{
  \label{fig:bccTi_w_disp}
  Comparison between the VIP method and the volume-varied SCPh calculations for bcc Ti.
  Vibrational atomic displacements were incorporated into the electronic free energy calculations.
  (a) The change in the Helmholtz free energy due to thermal expansion. 
  (b) The difference in the Gibbs free energy between the VIP method and the volume-varied SCPh calculations $G_{\rm SCPh}^{\rm volume}$.
  (c) The CLTE $\alpha_L$.
  The experimental data are taken from Ref.~\cite{Touloukian1975-tv}.
  The orange line is the standard VIP result ($B_1$ only, Eq.~(16)); the red line is a reference that also includes $B_2$ ($B_1+B_2$, Eqs.~(16) and (17)).
  The blue lines represent the results obtained by the volume-varied SCPh calculations.
  }
  \end{center}
\end{figure}

\begin{figure}[H]
  \begin{center}
  \includegraphics[width=1.0\linewidth]{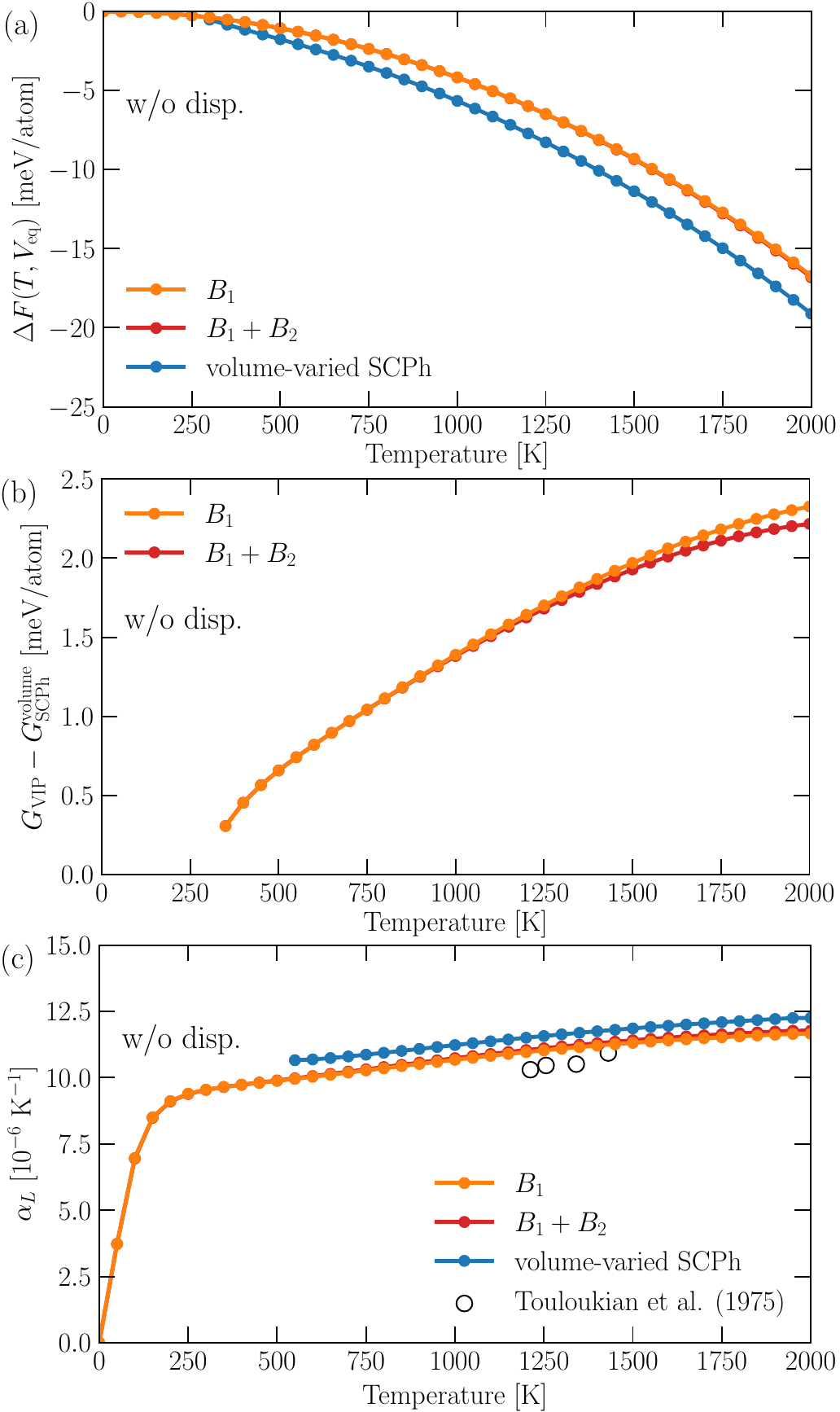}
  \caption{
  \label{fig:bccTi_wo_disp}
  Comparison between the VIP method and the volume-varied SCPh calculations for bcc Ti.
  Vibrational atomic displacements were not incorporated in the electronic free energy calculations.
  (a) The change in the Helmholtz free energy due to thermal expansion. 
  (b) The difference in the Gibbs free energy between the VIP method and the volume-varied SCPh calculations $G_{\rm SCPh}^{\rm volume}$.
  (c) The CLTE $\alpha_L$.
  The experimental data are taken from Ref.~\cite{Touloukian1975-tv}.
  The orange line is the standard VIP result ($B_1$ only, Eq.~(16)); the red line is a reference that also includes $B_2$ ($B_1+B_2$, Eqs.~(16) and (17)).
  The blue lines represent the results obtained by the volume-varied SCPh calculations.
  }
  \end{center}
\end{figure}

The comparisons between the VIP method and the volume-varied SCPh calculations for bcc Ti in terms of the free energy and the CLTE are shown in Figs.~\ref{fig:bccTi_w_disp} and \ref{fig:bccTi_wo_disp}.
To incorporate thermal expansion, we performed SCPh calculations at multiple volumes.
Figure~\ref{fig:bccTi_w_disp} considers vibrational atomic displacements in the electronic free energy; Fig.~\ref{fig:bccTi_wo_disp} does not.
$\\$
$\\$
\twocolumngrid

\newpage

\clearpage
\onecolumngrid
\subsubsection{hcp Ti}
\begin{figure}[H]
  \begin{center}
  \includegraphics[width=1.0\linewidth]{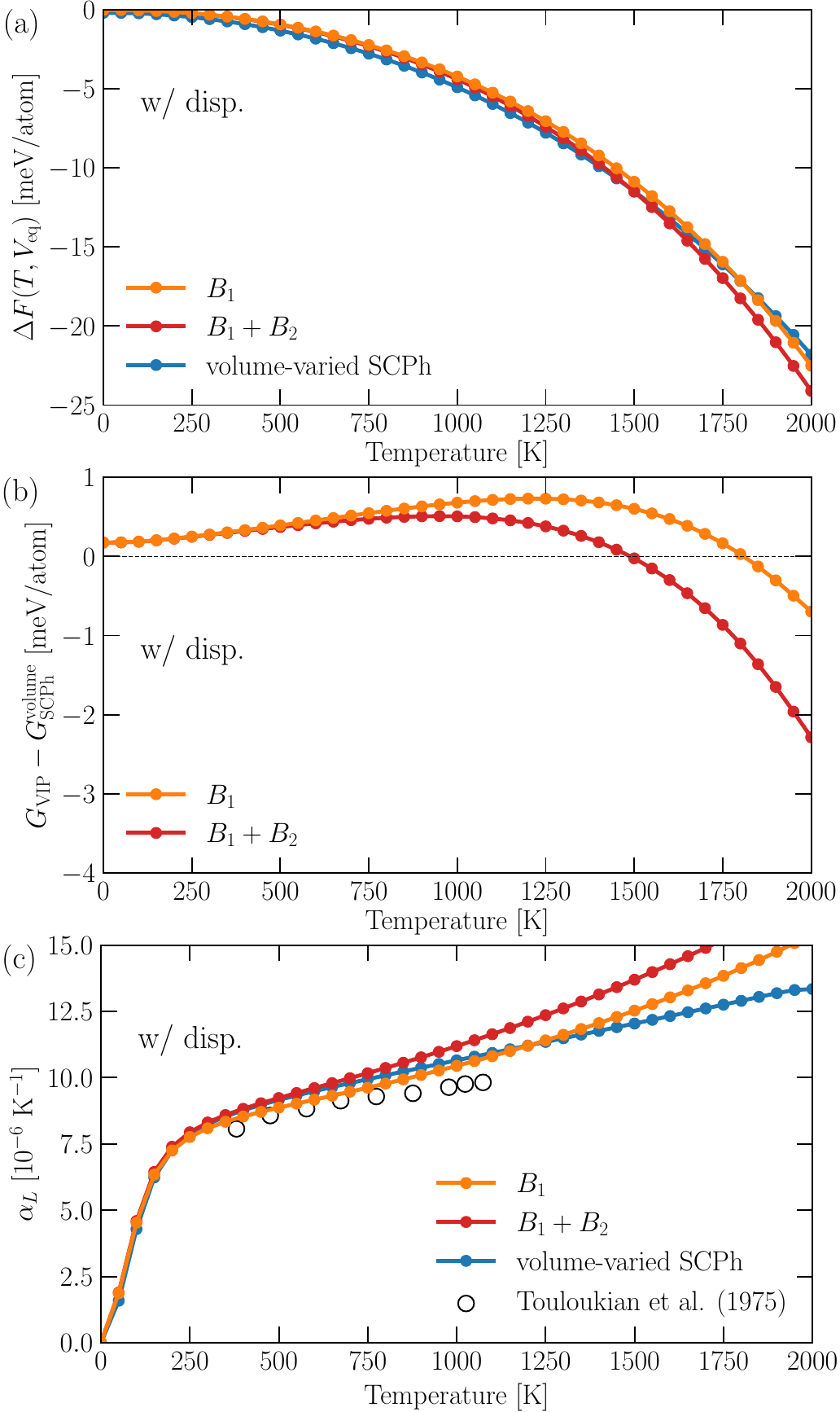}
  \caption{
  \label{fig:hcpTi_w_disp}
  Comparison between the VIP method and the volum-varied SCPh calculations for hcp Ti.
  Vibrational atomic displacements were incorporated into the electronic free energy calculations.
  (a) The change in the Helmholtz free energy due to thermal expansion. 
  (b) The difference in the Gibbs free energy between the VIP method and the volume-varied SCPh calculations $G_{\rm SCPh}^{\rm volume}$.
  (c) The CLTE $\alpha_L$.
  The experimental data are taken from Ref.~\cite{Touloukian1975-tv}.
  The orange line is the standard VIP result ($B_1$ only, Eq.~(16)); the red line is a reference that also includes $B_2$ ($B_1+B_2$, Eqs.~(16) and (17)).
  The blue lines represent the results obtained by the volume-varied SCPh calculations.
  }
  \end{center}
\end{figure}

\begin{figure}[H]
  \begin{center}
  \includegraphics[width=1.0\linewidth]{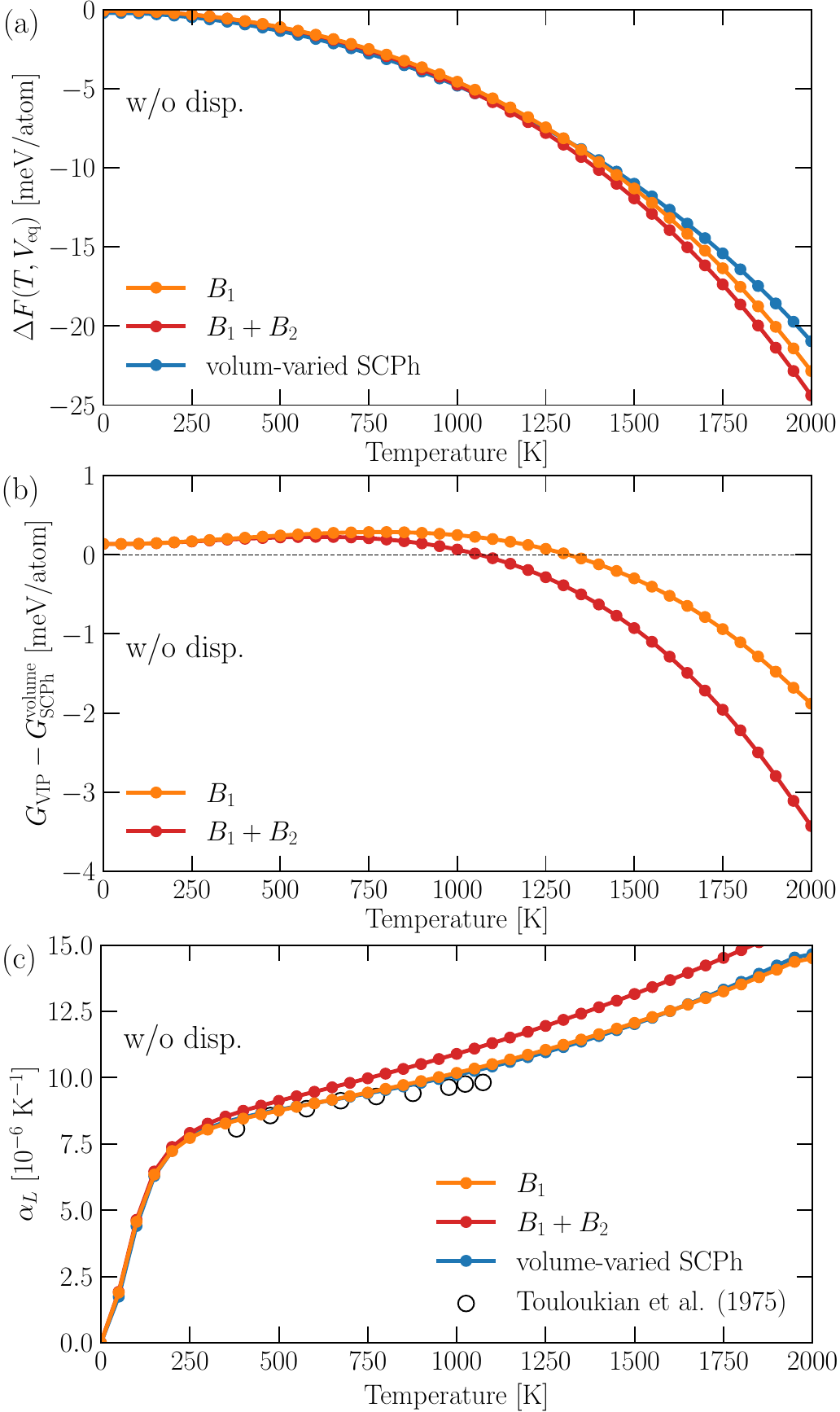}
  \caption{
  \label{fig:hcpTi_wo_disp}
  Comparison between the VIP method and the volume-varied SCPh calculations for hcp Ti.
  Vibrational atomic displacements were not incorporated in the electronic free energy calculations.
  (a) The change in the Helmholtz free energy due to thermal expansion. 
  (b) The difference in the Gibbs free energy between the VIP method and the volume-varied SCPh calculations $G_{\rm SCPh}^{\rm volume}$.
  (c) The CLTE $\alpha_L$.
  The experimental data are taken from Ref.~\cite{Touloukian1975-tv}.
  The orange line is the standard VIP result ($B_1$ only, Eq.~(16)); the red line is a reference that also includes $B_2$ ($B_1+B_2$, Eqs.~(16) and (17)).
  The blue lines represent the results obtained by the volume-varied SCPh calculations.
  }
  \end{center}
\end{figure}

The comparisons between the VIP method and the volume-varied SCPh calculations for hcp Ti in terms of the free energy and the CLTE are shown in Figs.~\ref{fig:hcpTi_w_disp} and \ref{fig:hcpTi_wo_disp}.
To incorporate thermal expansion, we performed multiple SCPh calculations at different volumes.
Figure~\ref{fig:hcpTi_w_disp} considers vibrational atomic displacements in the electronic free energy; Fig.~\ref{fig:hcpTi_wo_disp} does not.
$\\$
$\\$
\twocolumngrid

\newpage

\clearpage
\onecolumngrid
\subsection{tetragonal ZrO\texorpdfstring{$_2$}{2}}
\begin{figure}[H]
  \begin{center}
  \includegraphics[width=0.6\linewidth]{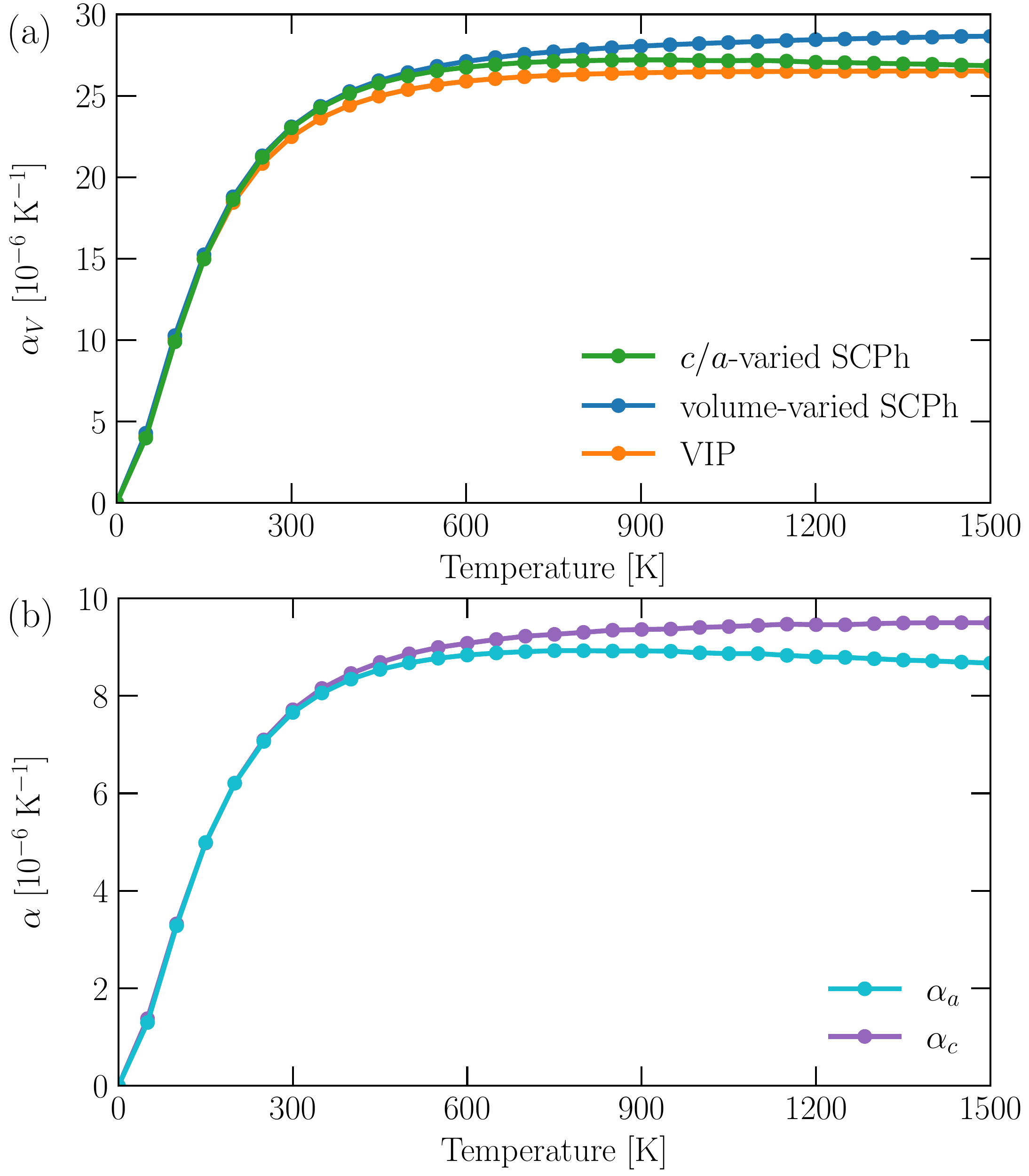}
  \caption{
  \label{fig:t-ZrO2_suppl}
  Coefficient of thermal expansions calculated by the $c/a$-varied SCPh calculations for tetragonal ZrO$_2$ from 0 to 1500 K. 
  (a) The CVTE $\alpha_V$.
  (b) The CLTE, $\alpha_a=1/a(da/dT)$ and $\alpha_c=1/c(dc/dT)$. 
  The green, blue, and orange lines represent the results obtained by the $c/a$-varied SCPh calculations, the volume-varied SCPh calculations, and the VIP method using $B_1$ in Eq.~(16), respectively.
  }
  \end{center}
\end{figure}

Figure~\ref{fig:t-ZrO2_suppl} shows the coefficient of thermal expansions for tetragonal ZrO$_2$ obtained by the $c/a$-varied SCPh calculations, which independently varied the $a$ and $c$ axes to minimize the Gibbs free energy at each temperature.
The resulting $c/a$ ratio of 1.443, exhibiting good agreement with the experimental result at 973 K (a deviation of only about 0.02 \%) presented in \cite{Igawa2001-ga}.
Furthermore, the ratio can be treated as constant in the temperature range of 0 to 1500 K, given that its maximum change from 0 K is less than 0.06 \%.
The coefficient of volumetric thermal expansion (CVTE) calculated from the volume-varied SCPh, agrees about 5 \% of the result obtained by the $c/a$-varied SCPh at 1200 K as shown in Fig.~\ref{fig:t-ZrO2_suppl} (a).
In addition, the Gibbs free energies calculated from the volume-varied SCPh reproduce those of $c/a$-varied SCPh calculations, with the deviation between them reaching a maximum of 0.4 meV/atom at 1500 K and lessening at lower temperatures.
Therefore, the volume-varied SCPh calculations can be utilized to accurately determine the Gibbs free energy for tetragonal ZrO$_2$, even without explicitly varying the $c/a$ ratio.
For reference, Fig.~\ref{fig:t-ZrO2_suppl} (b) also shows the CLTEs for the $a$ and $c$ axes.

